\documentclass[manuscript, anonymous=false]{acmart}
\AtBeginDocument{%
  }
\usepackage{multirow} 
\usepackage{graphicx}
\usepackage{subcaption}
\usepackage{url}
\usepackage{hyperref}
\setcopyright{acmcopyright}
\copyrightyear{2024}
\acmYear{2024}
\acmDOI{XXXXXXX.XXXXXXX}

\acmJournal{TORS}

\begin{document}

\title{Consistency Regularization for Complementary Clothing Recommendations}

		
		
		
		
		

\author{Shuiying Liao}
\email{21122587r@connect.polyu.hk}
\orcid{0000-0002-8331-6486}
\affiliation{
  \institution{The Hong Kong Polytechnic University}
  \country{Hong Kong}
}

\author{P.Y. Mok}
    \email{tracy.mok@polyu.edu.hk}
    \orcid{0000-0002-0635-5318}
   \affiliation{
        \institution{The Hong Kong Polytechnic University}
        \country{Hong Kong}
 }
\thanks{Co-corresponding author: tracy.mok@polyu.edu.hk}
 
\author{Li Li}
    \email{lillyli@ust.hk}
    \orcid{0000-0003-0622-4497}
   \affiliation{
        \institution{The Hong Kong University of Science and Technology}
        \country{Hong Kong}
}
\thanks{Co-corresponding author: lillyli@ust.hk}


\begin{abstract}
 
This paper reports on the development of a Consistency Regularized model for Bayesian Personalized Ranking (CR-BPR), addressing to the drawbacks in existing complementary clothing recommendation methods, namely limited consistency and biased learning caused by diverse feature scale of multi-modal data. Compared to other product types, fashion preferences are inherently subjective and more personal, and fashion are often presented, not by individual clothing product, but with other complementary product(s) in a well coordinated fashion outfit.  Current complementary-product recommendation studies primarily focus on user preference and product matching, this study further emphasizes the \textit{consistency} observed in user-product interactions as well as product-product interactions, in the specific context of clothing matching. Most traditional approaches often underplayed the impact of existing wardrobe items on future matching choices, resulting in less effective preference prediction models. Moreover, many multi-modal information based models overlook the limitations arising from various feature scales being involved. 
To address these research gaps, the CR-BPR model integrates collaborative filtering techniques to incorporate both user preference and product matching modeling, with a unique focus on consistency regularization for each aspect. The model not only learns user preferences and product compatibility but also maintains consistency in these interactions. Additionally, the incorporation of a \textit{feature scaling} process further addresses the imbalances caused by different feature scales, ensuring that the model can effectively handle multi-modal data without being skewed by any particular type of feature.
The effectiveness of the CR-BPR model was validated through detailed analysis involving two well-known clothing matching recommendation benchmark datasets. The results confirmed that the proposed approach significantly outperforms existing models in terms of prediction accuracy. The CR-BPR model presents a significant advancement in the development of personalized clothing matching recommendation, and provides a more comprehensive and effective solution for predicting user preferences and enhancing the overall user experience in fashion shopping e-commerce.
\end{abstract}

\maketitle


  
\section{Introduction}
Fashion is an important component of the global economy and also an essential aspect of human culture and life. A major problem associated with fashion is how best to select and match clothes when presented with a large number of choices, this being particularly difficult for people who lack the necessary fashion knowledge. To assist in this respect, the data-driven clothing matching approach has recently been receiving attention, particularly in the light of computational advances~\cite{guan2016apparel, song2017neurostylist, GPBPR, han2019prototype, li2019semi, gao2019fashion, ding2021leveraging, 9451610}. 
This problem is challenging, however, due to the abundance of different fashion products, and the performance of existing models remains inadequate. Moreover, since fashionable clothing matching is subjective, i.e., different people prefer different matching styles, it is essential to take personal preferences into consideration to achieve effective personalized clothing matching, which requires the modeling of two key aspects: user preference and product matching.
More specifically, \textit{user preference} considers and measures how appealing a clothing product is for a particular user. \textit{Product matching}, on the other hand, considers and measures how well two products are complementary to each other as a harmony and compatible outfit. 
To this end, existing methods of collaborative filtering (CF) based frameworks have two branches with different focuses: one for modeling user-product relationships and the other for modeling product-product relationships. For instance, studies such as~\cite{GPBPR, PAI, PCE} have implemented this dual-branch approach, leveraging collaborative filtering techniques to simultaneously model these two relationships, and leading to more accurate and personalized clothing matching recommendations.

Nevertheless, modeling user preference and product matching simultaneously is difficult due to the coupling effect and the complexity of each underlying pattern. In practice, data available for the learning of the two patterns are inadequate. For example, the interaction data of user and product pairs are sparse, whereas both the user set and product set are often of large number. Such inadequate interaction data may create more difficulties in terms of the problem being investigated. To tackle the \emph{data sparsity} problem, most approaches to date have attempted to improve on the modeling of the two respective patterns. For example, attribute information of clothing products is leveraged to model fine-grained user preferences on products and product matching (pairwise clothing) patterns~\cite{PAI}. Incorporating such additional information has also been applied in other recommendation related studies~\cite{ ma2024personalized}. Despite their effectiveness, such additional information generally requires additional efforts, manually or computationally, which introduces extra cost and/or data noise.

\begin{figure}[t]
  \centering
  \includegraphics[width=\linewidth]{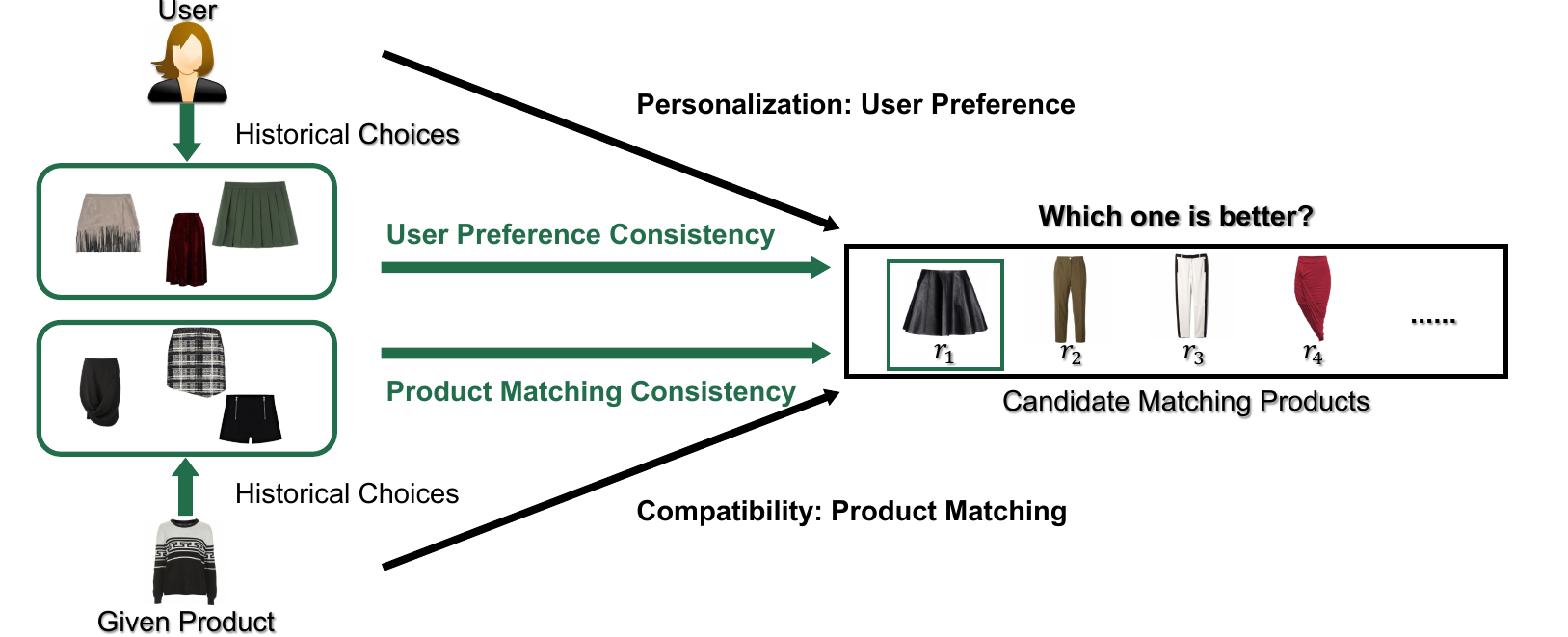}
  \caption{Key concepts of personalized clothing matching}
  \label{fig:1}
\end{figure}

Common sense shows that there is \textit{consistency} in both user preference and clothing matching patterns~\cite{he2020consistency, kim2009applying, style_consistency}, although this has been largely ignored in most previous studies. For example, online fashion customers tend to shop for products with the same brand, style, or color~\cite{POG}. As illustrated in Fig.~\ref{fig:1}, when presented with multiple choices, the user (\textit{$u$}) may prefer clothing similar to what they have chosen before. Similarly, from the perspective of clothing matching, given a top garment of the sweater (\textit{$g$}) in Fig.~\ref{fig:1}, choosing a bottom clothing similar to what \textit{$g$} has matched before would be a reasonable recommendation. Considering both aspects simultaneously indicates that the black skirt (\textit{$r_1$}) would be the choice from the point of view of consistency. In some real-world situations, the consistency of a user's behavior may weigh more heavily than fitting in with other users~\cite{kim2009applying}. This encourages incorporating consistency in the model so as to improve personalized clothing matching. On one hand, consistency modeling enables more information to be leveraged than only the user-product interactions for recommending complementary product that fit user's personal preferences and coordinate well with the given product. On the other hand, consistency branch could serve as a regularization means for the CF-based methods, which are mostly based upon exploring the similarity of user/product interaction patterns. 

In addition to the above, since such models predict the user preference and product matching scores directly based on features, the prediction can be affected by the scales of various features being used. More importantly, the majority of the personalized clothing matching models project recommendation scores based on more than one type of feature, which naturally vary in scale since they originate from different data sources. Multi-modal features involving different scales thus contribute to the recommendation scores at different levels, thereby introducing errors due to incorrectly assigning the relative importance of the various features. In spite of this, such \textit{feature scale} related issues have been ignored in most existing models~\cite{GPBPR, PAI, computer, VBPR}, thereby adversely impacting their overall performance.

Based upon the above, this paper proposes a novel \textbf{C}onsistency \textbf{R}egularized \textbf{B}ayesian \textbf{P}ersonalized \textbf{R}anking (\textbf{CR-BPR}) model with \textbf{\textit{Feature Scaling}}. The proposed \textbf{CR-BPR} is based on dual BPR that models user preference and product matching simultaneously. Each BPR branch is based on a hybrid collaborative filtering module, with latent and multi-modal feature inputs, which have proven to be an effective baseline in previous studies~\cite{GPBPR}. In order to tackle the relative feature importance problem resulting from differences in feature scales, a feature scaling process is introduced into the CR-BPR model. Specifically, different types of features, including multi-modal content features and latent features, are processed with a purpose-specific normalization transformation before being used for score prediction for user preference and product matching, respectively. Furthermore, to facilitate \textit{consistency} modeling, two branches for consistency regularization are designed, which specifically measure the \textit{similarity} between the target product and historical choices from both user preference and product matching perspectives. The two branches of consistency modeling are integrated with the basic user preference and product matching modeling into a multi-branch BPR model being optimized jointly. 

The main contributions of this paper are summarized as follows:
\begin{itemize}
\item A new CR-BPR model is developed, based on a dual BPR model, which simultaneously addresses user preference and product matching in clothing recommendation. In particular, product consistency is leveraged from both the user preference and product matching perspectives to regulate the modeling of the two key aspects and further boost overall performance.
\item The proposed CR-BPR also incorporates a feature scaling function to tackle the relative feature importance problem present in most models with multi-modal data features, which has been largely overlooked in previous studies.
\item The newly developed CR-BPR was validated by means of detailed analysis based on two benchmark fashion recommendation datasets: IQON3000 and Polyvore-519. This demonstrated the state-of-the-art performance of the newly developed model.
\end{itemize}

The rest of the article is organized as follows: Section~\ref{sec_2}  provides a brief review of related work. In Section~\ref{sec_3}, we elaborate on the proposed CR-BPR model. Section \ref{sec_4} presents the experimental results and in-depth discussion. Finally, Section~\ref{sec_5} concludes the paper and discusses potential directions for future research.

\section{Related work}\label{sec_2}
\subsection{Compatibility Modeling}
Clothing matching studies have garnered significant attention due to their practical relevance in daily life and commercial applications.
\textbf{Compatibility} is the most important criterion in clothing matching to assess whether various clothing products fit well together.  
Initial studies~\cite{vasileva2018learning, yang2020learning, dong2020fashion, kaicheng2021modeling, jing2019low} primarily focused on capturing matching compatibility that are visually compatible or functionally complementary, yielding validated results. 
Visual features, such as color, texture, and style, provide essential data for determining the compatibility of clothing products.
By extracting these features using techniques like CNNs, researchers can mathematically assess compatibility based on the distance between item vectors in the latent space.
For example, Li \emph{et al.}~\cite{li2021attribute} and Feng \emph{et al.}~\cite{part_embedding} have made significant contributions in attribute-level matching by using visual features and constructing complex graphs for outfit composition. To further captured functional complementary, Jing \emph{et al.}~\cite{tmm2022} introduced a tripartite graph model that encodes correlations among features, clothing products, and outfits, thereby providing a richer representation of matching relationships. 
\textit{However, they often overlooked the nuanced preferences of individual users, leading to generic recommendations that might less effective in personalized recommendations.}

\subsection{Personalization Modeling}
Benefiting from on-going computational improvements, studies~\cite{trakulwaranont2022personalized, ding2021leveraging, velivckovic2017graph} have proved that user-personalized data can greatly improve prediction accuracy from a \textbf{user-personalized} perspective. 
Without loss of generality, advanced personalized compatibility modeling methods focus on both the given product and its matching product while considering the user-specific preferences.
To this end, GP-BPR~\cite{GPBPR} firstly used both user-product interactions for personal preference modeling and product-product interactions for product matching modeling from a general aesthetic perspective.
Other fine-grained preference attention modules~\cite{hou2019explainable, A3, computer} are also widely explored that can describe the reasons for recommending clothing in a personalized way by semantic highlighting visual attributes. 
Another category of methods attempted to improve personalized recommendation accuracy by manually acquiring additional product attributes~\cite{PAI} or user information \cite{wang2017item, ma2024personalized} as supplementary content, whereas the cost of data mining for such information is often prohibitive, restricting these methods from being generalized across different datasets.
To achieve higher performance for personalized clothing matching recommendation, advanced efforts have introduced multi-modal information. However, various pre-trained models, such as ResNet\cite{resnet50}, commonly used in the general deep learning domain, primarily extract features for classification tasks rather than recommendation systems. In recommendation systems, it is crucial to leverage the similarity of these multi-modal features, often measured through methods like inner product calculations. The magnitude of various feature vectors from different source might introduce imbalanced latent representation learning, and further influence the similarity calculation, \textit{but which has been largely overlooked in previous personalized clothing matching recommendations.}

\subsection{Regularizations for Data-drive Recommendations}
\begin{figure*}[th]
  \centering
  \includegraphics[width=\textwidth]{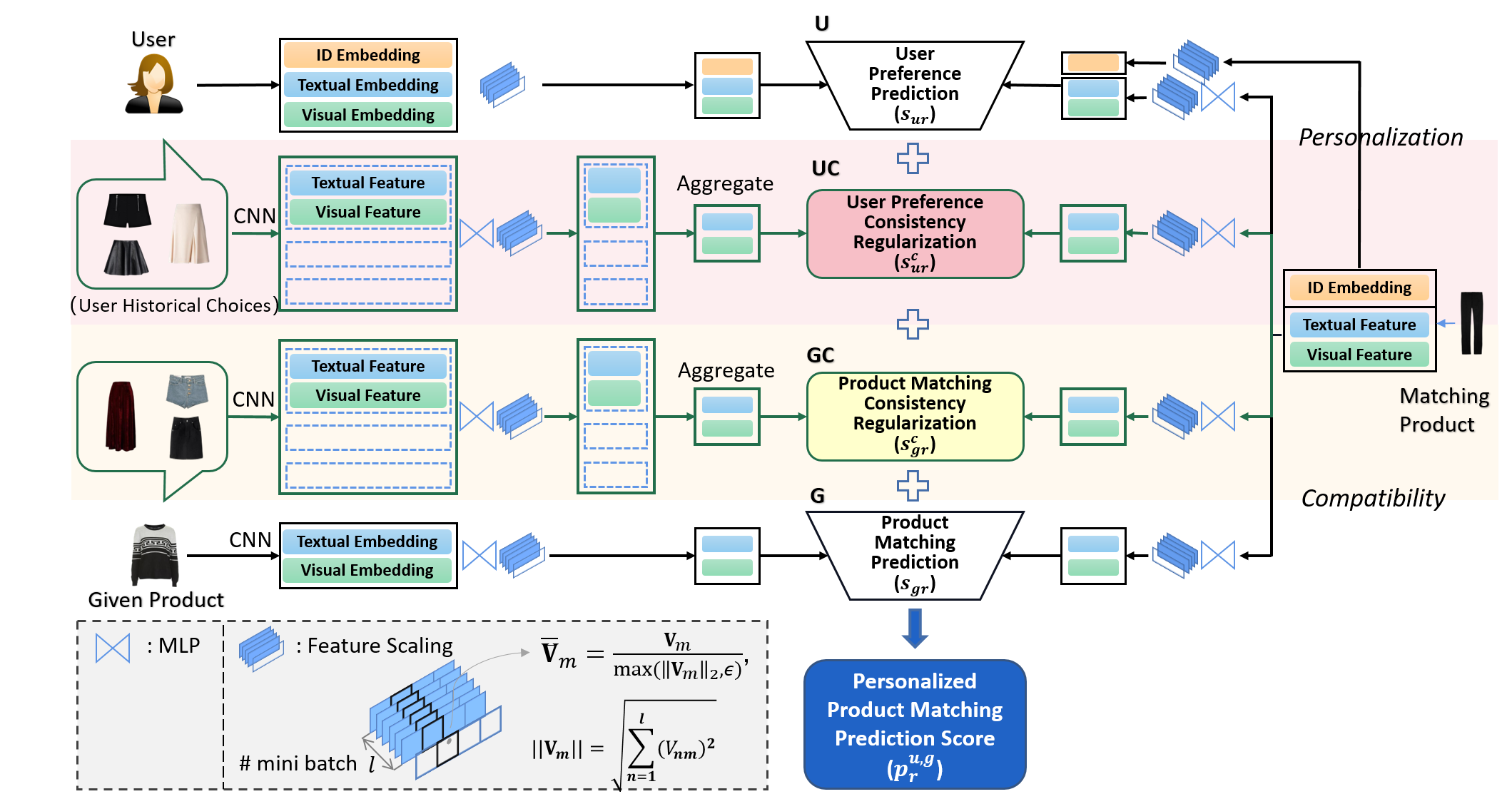}
  \caption{The CR-BPR scheme has four main components. User preferences are predicted by user-product interactions. Product Matching Prediction module uses visual and textual product-product interaction principles. The User Preference Consistency regularization module captures the average similarity between the target matching garment and user historical choices on both visual and textual modalities. The Product Matching Consistency regularization module determines the mean similarity between the target matching garment and the historical choices matched for the given clothing.}
  \label{fig:method}
\end{figure*}

Additionally, the performance of recommendation models usually be hampered by data problems such as the presence of noise and bias \cite{liu2024dual}, data sparsity \cite{zhao2024hierarchical, xu2020neural}, and cold start \cite{wang2023multifaceted}. Effectively leveraging existing data to provide accurate personalized recommendations has always been a key focus in recommendation systems. 
Previous approaches have provided comprehensive solutions to address these methods, such as feature-aware matrix factorization~\cite{chen2016context, VBPR, GPBPR, liao, PAI}, multi-modal information connected GCNs~\cite{sun2020multi, he2021click, lei2023learning}, or cross domain recommendation strategies~\cite{liu2024privacy, zhu2019dtcdr, zhu2019dtcdr}.
However, these techniques are typically less effective or computationally too expensive. 
Additionally, besides the aspect of user preference and general product matching,
the matching consistent relationship present in the interaction data has not been thoroughly investigated. 
Different from previous methods, we use a simple but effective regularization methods to overcome the data sparsity issue. 
In general definition,
\textbf{regularization} often refers to techniques being used to calibrate data-driven models to avoid overfitting or underfitting. Typical regularization techniques include L1~\cite{tibshirani1996regression} and L2~\cite{hoerl1970ridge} regularization, and various techniques have been successfully applied in many applications, such as matrix factorization \cite{tran2018regularizing, liang2016factorization, liu2021bayesian} and implicit feedback \cite{hu2008collaborative, BPR, sheth2023causal}. Modeling the consistency present in interaction data can be viewed as another way of regularization or regulating. 
\textit{This provides the basis and motivation for the present CR-BPR approach, namely comprehensively model product matching and user preference with regularization by means of implicit consistency relationships (see Fig.~\ref{fig:method}).} 



\section{Method}\label{sec_3}
The proposed CR-BPR method characterizes the personalized clothing matching problem into four main parts, including (1) user preference modeling, (2) product matching modeling, (3) user preference consistency modeling, and (4) product matching consistency modeling. All four parts are based on a normalized representation module (i.e., feature scaling),  which will be introduced in the section immediately after the section for `Problem Formulation'. The four modeling parts will then be explained, while the optimization of the whole CR-BPR model will be discussed at the end.

\subsection{Problem Formulation}
Here the personalized clothing matching problem is addressed by recommending to each user a fashion garment that matches well with a given piece of clothing. For example, the recommendation for the bottom clothing is given to a user for a specified top garment as a clothing outfit, and vice versa. 
Let $\mathcal{U}=\left\{u_1, u_2, \ldots, u_{|\mathcal{U}|}\right\}$, $\mathcal{G}=\left\{g_1, g_2, \ldots, g_{|\mathcal{G}|}\right\}$ and $\mathcal{R}=\left\{r_1, r_2, \ldots, r_{|\mathcal{R}|}\right\}$ represent the sets of users, given products and recommended matching products, respectively. For simplicity, the subscript is omitted and $u$, $g$, and $r$ are used to represent a specific user, given product, and recommended matching product, respectively, in the rest of this paper. An outfit $<g, r>$ is simply defined as the combination of one given product $g \in \mathcal{G}$ and one recommended matching product $r \in \mathcal{R}$. Each user $u\in{\mathcal{U}}$ interacts with several product pairs, resulting in a set of triplets $<u, g, r>$ that represents the interaction records of user $u$, with multi-modal features being adopted to describe clothing products $<g, r>$. For example, the given garment $g$ is represented by its visual feature $\mathbf{v}_g \in \mathbb{R}^{d_v}$ and textual feature $\mathbf{w}_g\in \mathbb{R}^{d_w}$, where $d_v$ and $d_w$ denote the dimensions of the two features. Similarly, the visual and textual features for the recommended matching product $r$ are $\mathbf{v}_r$ and $\mathbf{w}_r$. This study aims to develop a recommendation model $\mathcal{F}$ that can effectively predict the preference score between any triplet of user-product-recommended matching 
$<u, g, r>$:
\begin{equation}
p^{u,g}_{r} = \mathcal{F}(u, g, r | \Theta),
\label{eq:overall_obj}
\end{equation}
where $\Theta$ denotes all parameters in $\mathcal{F}$. The score $p^{u,g}_{r}$ represents the possibility of the matching product $r$ being preferred by the user $u$ and still matching well with the given product $g$. An effective model $\mathcal{F}$ should be able to score the compatible triplets higher than the incompatible ones. 

\subsection{Representations with Feature Scaling}
\label{subsec:item_rep}
The representations of a clothing product in the fashion matching problem are usually based on multi-modal features extracted from pre-trained deep neural networks, such as CNN, whereas the visual and textual features are included in this work. Advances in computer vision and neural language processing enable high-quality features from product images and textual descriptions to be derived. These two modalities of information can generally cover the main characteristics of clothing products. Nevertheless, such product features are extracted from neural network models pre-trained for other different tasks, such as image classification. Therefore, to make them applicable for the modeling of user preferences or for other study objectives, a non-linear transformation is applied on top of the extracted product features. 
Specifically, different multi-layer perceptrons (MLPs) are applied to process product features to obtain different representations used in different parts of the whole framework, as shown in Fig.~\ref{fig:method}. 
Different MLPs are used to obtain modality-specific representations for product features of different modalities. Take visual modality as an example, the pre-trained features $\mathbf{v}_r$ (for recommended matching product $r$) go through a stack of linear projection and non-linear activation layers transformed into visual representations as follows:
\begin{equation}
\left\{
\begin{aligned}
&\mathbf{v}^1_r=\sigma\left(\mathbf{W}^1 \mathbf{v}_r+\mathbf{b}^1\right) \\
&\mathbf{v}^k_r=\sigma\left(\mathbf{W}^k\mathbf{v}_r^{k-1}+\mathbf{b}^k\right), k=2, \cdots, K,
\end{aligned}
\right.
\label{eq.mlp}
\end{equation}
where $\mathbf{W}^k$ and $\mathbf{b}^k$ are the projection matrix and bias vector in the $k$-th layer, respectively, $K$ represents the total number of layers in the MLP network, and the K-th layer refers to the final output layer. $\sigma$ is the activation function, which is implemented by means of the Sigmoid function. For simplicity, $\text{MLP}_v$ is used to denote the feature processing module described in eq. (\ref{eq.mlp}), in which subscript $v$ refers specifically to the visual features. The resulting new visual representation from the output layer is: 
\begin{equation}
\mathbf{v}^*_r = \mathbf{v}^K_r = \text{MLP}_{v}(\mathbf{v}_r) \in \mathbb{R}^{d^*_v}, 
\label{eq:output_feature}
\end{equation}
where $\mathbb{R}^{d^*_v}$ is the corresponding dimension. 
The textual features are subject to similar procedures, in another module $\text{MLP}_w$, to obtain enhanced textural representations. Specifically, for the recommended matching product $r$, the textural representation is $\mathbf{w}^*_r \in \mathbb{R}^{d^*_w}$.

To address the issue caused by different feature scales that may impact preference modeling, batch feature scaling is applied to features before modeling user-product and product-product interactions. Taking the visual representation as an example, given one batch of matching products, the visual representation tensor can be denoted as $\mathbf{V} \in \mathbb{R}^{d_v \times l}$, where $l$ represents the batch size. The feature scaling along the dimension $m \in [0,1,...,d_v]$: $\mathbf{V}_{m} \in \mathbb{R}^{1 \times l}$ is transformed by:
\begin{subequations}
\label{eq:norm}
\begin{eqnarray}
\mathbf{\bar{V}}_{m} = \frac{\mathbf{V}_{m}}{\text{max}(||\mathbf{V}_{m}||_2, \epsilon)}, ~~~~~~~~~~\\
\textrm{where} \qquad \qquad \qquad ||\mathbf{V}_{m}||_2 = \sqrt{\sum_{n=1}^{l}(V_{nm})^2}~~~~~~~~~ 
\end{eqnarray}
\end{subequations}
and $\epsilon$=1$e-$12. Therefore, after the feature scaling, the batch representations transform into $\mathbf{\bar{V}} = [\mathbf{\bar{V}}_{0}^T, \mathbf{\bar{V}}_{1}^T,...,\mathbf{\bar{V}}_{m}^T,$ $ ..., \mathbf{\bar{V}}_{d_v}^T]^T \in \mathbb{R}^{d_v \times l}$. Thereafter, the specific representation of matching item $r$, namely $\mathbf{\bar{v}}_r$, can be obtained. For simplicity, the feature scaling process (eq.~(\ref{eq:norm})) is abbreviated as $\text{Norm}(\cdot)$, giving $\mathbf{\bar{v}_r}=\text{Norm}(\mathbf{v}^*_r)$. Similarly, the textual representation is also normalized as $\mathbf{\bar{w}}_r=\text{Norm}(\mathbf{w}^*_r)$.

\subsection{User Preference and Product Matching Modeling}
\subsubsection{User preference modeling (\textbf{U})}
The user preference predictor is based on matrix factorization (MF)~\cite{BPR, VBPR}, which has remarkable success in various personalized recommendation tasks. The underlying philosophy is to break down the user-product interaction matrix into user latent and product latent factors, the inner product of which encodes an estimate of the user-product interaction. To effectively incorporate the multi-modal product information, the user preference for a specific product content is also explored. Based on the Visual Bayesian Personalized Ranking (V-BPR) model, the following user preference predictor is designed:
\begin{equation}
s_{u r}=(\mathbf{\bar{e}}_u)^T \mathbf{\bar{e}}_r+\eta\left(\mathbf{\bar{v}}_u\right)^T \mathbf{\bar{v}}_r^p+(1-\eta)\left(\mathbf{\bar{w}}_u\right)^T \mathbf{\bar{w}}_r^p+\beta_u+\beta_r+ \alpha 
\label{eq.sub}
\end{equation}
where $\mathbf{\bar{e}}_u$ and $\mathbf{\bar{e}}_r$ are normalized latent embeddings for user $u$ and product $r$ entities, respectively; $\mathbf{\bar{v}}_r^p$ and $\mathbf{\bar{w}}_r^p$ are multi-modal representations of $r$ obtained from the non-linear transformation (eqs. (\ref{eq.mlp}) and (\ref{eq:output_feature})) and feature scaling (eq. (\ref{eq:norm})) operations introduced in previous section~\ref{subsec:item_rep}. The subscript $p$ indicates the representations are exclusive to the user's latent preference, i.e., \textbf{\textit{p}}ersonalization modeling, obtained by applying exclusive MLPs as follows:
\begin{subequations}
\label{eq.norm-mlp}
\begin{eqnarray}
    \mathbf{\bar{v}}^p_r = \text{Norm}(\text{MLP}^p_v(\mathbf{v}_r)), \\
    \mathbf{\bar{w}}^p_r = \text{Norm}(\text{MLP}^p_w(\mathbf{w}_r)).
\end{eqnarray}
\end{subequations}
The embedding tables for user $\mathbf{E}^U \in \mathbb{R}^{|\mathcal{U}| \times d_e}$ and item $\mathbf{E}^I \in \mathbb{R}^{|(\mathcal{G}+\mathcal{R})| \times d_e}$ are initialized. 
For a given user $u$ (or matching product $r$), the latent embedding $\mathbf{e}_u \in \mathbf{E}^U$ ($\mathbf{e}_r \in \mathbf{E}^I$) is normalized as $\mathbf{\bar{e}}_u = \text{Norm}(\mathbf{e}_u)$ ($\mathbf{\bar{e}}_r = \text{Norm}(\mathbf{e}_r)$) before being used for user preference modeling. Such an operation is applied to align the scale of different middle-stage representations, and which will be evaluated in terms of its improvement of the overall performance.
$\mathbf{\bar{v}}_u$ and $\mathbf{\bar{w}}_u$ in eq. (\ref{eq.sub}) are another two user representations associated with the user ID for the modeling of the implicit user visual and textural preferences, which are also normalized after embedding table initialization, as $\mathbf{\bar{e}}_u$ does in eq. (\ref{eq.norm-mlp}). All preference scores are measured with the inner product operation involving the corresponding user and matching product representations.   
Moreover, in eq. (\ref{eq.sub}), $\alpha$ is the global offset, $\beta_u$ and $\beta_r$ are user and product bias terms, which are standard in the score function of MF-BPR models. The trade-off parameter $\eta$ is applied to balance the relative importance of visual and textural preferences~\cite{GPBPR}, at the same time ensuring the latent and content user preferences contribute equally to the overall user preference score prediction.

\subsubsection{Product matching modeling (\textbf{G})}
The product matching modeling predicts the matching relationship between a pair of products from a product matching perspective. By modeling product-product interactions based on content information, potential matching cues from some specific factors about fashion products, such as color, texture, style and function ~\cite{sun2020learning} are explored. The present matching relationship modeling is based on the visual and textual product matching of the two clothing items. With a given clothing product $g$ and a matching product  $r$, their normalized visual representations are $\mathbf{\bar{v}}^m_g$ and $\mathbf{\bar{v}}^m_r$, and textual representations are $\mathbf{\bar{w}}^m_g$ and $\mathbf{\bar{w}}^m_r$, respectively. The subscript $m$ denotes that the representations are exclusive to \textit{\textbf{matching}} modeling, as explained in Eq.(\ref{eq.norm-mlp}). Specifically, the visual and textual representations of a given garment $g$ are $\mathbf{\bar{v}}^m_g = \text{Norm}(\text{MLP}^m_v(\mathbf{v}_g))$ and $\mathbf{\bar{w}}^m_g = \text{Norm}(\text{MLP}^m_w(\mathbf{w}_g))$, respectively. 
The dot product is applied to representation vectors in order to model the matching relationships between two products in each modality. The visual and textual matching results are integrated with a linear combination to obtain the multi-modal matching score between the two clothing products. This is defined as a multi-modal product matching predictor $s_{gr}$ as follows:
\begin{equation}
\begin{aligned}
s_{g r}=\pi\left(\mathbf{\bar{v}}^m_g\right)^T \mathbf{\bar{v}}_r^m+(1-\pi)\left(\mathbf{\bar{w}}^m_g\right)^T \mathbf{\bar{w}}_r^m, 
\label{eq.stb}
\end{aligned}
\end{equation}
where $\pi$ is a trade-off parameter to balance the importance of visual and textual modalities in product matching modeling. The $s_{g r}$ value captures overall product matching score between a given garment $g$ and a matching product $r$.

\subsection{Consistency Regularization Modeling}
\label{subset_Con}
This section introduces the consistency regularization module, which includes two branches covering both the user preference and product matching aspects. Specifically, two consistency regularization modules, namely the \textit{user preference consistency} and \textit{proudct matching consistency} regularization modules are designed.

\subsubsection{User preference consistency regularization (\textbf{UC})}
To measure user preference consistency, it is proposed to measure the average similarity between a target matching product $r$ (e.g., the bottom clothing) and historical choices of the user $u$, namely the bottom clothing products the user previously selected or reviewed recorded in user-product interaction data. A higher similarity score means a stronger user preference consistency, and a greater possibility of the product being chosen as a match for the specified garment $g$. 

User preference consistency (UC) modeling faces many challenges, such as different lengths of the available user-product interaction records as well as the fact that user preferences may change dynamically. 
A classic solution~\cite{he2020consistency} is using the recent products to capture the current preference as a basis for a user-generated priority product list. 
Nevertheless, personalized clothing matching records may not always contain a time stamp revealing the order in which each user-product interaction occurs. 
To address these challenges, each target matching product interaction with user $u$ is first of all aggregated in order to produce an unsequenced list of historical choices $ut$, representing each user's overall preference. 
To detail the consistency priority of the current preference, the products with original features most similar to the target matching product are filtered out. 
For simplicity, the user-product historical choices are filtered as the $ur$ list in the rest of this article, while the original features refer to the pre-trained visual and textual features of each matching product. To reduce the effect of variable length bias of historical interaction, the length of the $ur$ lists for each user $u$ is kept the same and relatively short. The following regularization module shows how the user preference consistency is captured:
\begin{equation}
s_{u r}^c=\pi\left(\mathbf{\bar{v}}_{u r}\right)^T \mathbf{\bar{v}}_r^{c_u} +(1-\pi)\left(\mathbf{\bar{w}}_{u r}\right)^T \mathbf{\bar{w}}_r^{c_u}.\\
\label{eq.suc}
\end{equation}
For a target recommended matching product $r$, the normalized visual representations and textual representations are $\mathbf{\bar{v}}_r^{c_u} = \text{Norm}(\text{MLP}^{c_u}_v(\mathbf{v}_r))$ and $\mathbf{\bar{w}}^{c_u}_r = \text{Norm}(\text{MLP}^{c_u}_w(\mathbf{w}_r))$, respectively. The subscript $c_u$ denotes that the representations are exclusive to the modeling of \textit{\textbf{U}ser preference \textbf{C}onsistency}. $\mathbf{\bar{v}}_{u r}$ is an abbreviation of: 
\begin{equation}
\mathbf{\bar{v}}_{u r} = \frac{1}{N} \sum_{i=1}^N \mathbf{\bar{v}}_{u r}^{i},
\label{eq.N}
\end{equation}
where $N$ is the number of historical choices and $\mathbf{\bar{v}}_{u r}^{i}$ is the latent visual representation of the $i$-th product in the $ur$ list obtained from $\mathbf{\bar{v}}_{u r}^{i} = \text{Norm}(\text{MLP}^{c_u}_v(\mathbf{v}_{u r}))$. Similarly, there is a mean textual representation of $\mathbf{\bar{w}}_{u r}$ in the $ur$ list, which is calculated as in eq.(\ref{eq.N}).
Here, a content-based product similarity $s_{u r}^c$ is employed, which is calculated as dot product of the inherent characteristics of product pairs of target matching product $r$ and those in the $ur$ list of historical choices. The superscript $c$ in eq. (\ref{eq.suc}) is used so as to distinguish it from the user preference score $s_{u r}$ in the user preference modeling (eq. (\ref{eq.sub})). To simplify the weight adjustment, the same trade-off parameters $\pi$ and $1-\pi$ as in the other branches are used to represent the importance of visual and textual modality for the purpose of the similarity calculation in eq. (\ref{eq.suc}).


With the UC branch, the historical choices of user $u$ for target matching product $r$ will vary according to each triplet's similarity score. The value of the UC regularization module increases in line with the similarity score. Moreover, if user $u$ has limited user-product interaction, the predicted preference derived from the interaction information (eq. (\ref{eq.sub})) is inadequate to represent the user's actual choices~\cite{broadlearning}. The UC branch is used as auxiliary information to enable the model to provide better predictions.

\subsubsection{Product matching consistency regularization (\textbf{GC})} Similar to the UC branch modeling, another regularization module is employed to capture the product consistency from a product matching perspective. Similarly, all the matching products which interacted with specified garment $g$ are grouped and after which products that are most similar to the current target matching product $r$ are filtered based on pre-trained features. The filtered historical choices are simply referred as the $gr$ list, which characterizes the matching preference for the given product $g$. Thereafter, the average similarity between each target matching product $r$ and the products in the $gr$ list is calculated. Furthermore, $\mathbf{\bar{v}}_r^{c_g}$ and $\mathbf{\bar{w}}_{g r}^{c_g}$ are used to denote the normalized visual and textual representations of the specified garment $g$, obtained from $\mathbf{\bar{v}}_r^{c_g} = \text{Norm}(\text{MLP}^{c_g}_v(\mathbf{v}_r))$ and $\mathbf{\bar{w}}^{c_g}_r = \text{Norm}(\text{MLP}^{c_g}_w(\mathbf{w}_r))$, respectively. Subscript ${c_g}$ denotes the representations are exclusive for the purpose of \textit{product matching \textbf{C}onsistency} modeling. $\mathbf{\bar{v}}_{g r}$ is used to represent the mean visual aggregation of each target matching product in the $gr$ list. Similarly, given the number of the historical choice $N$, this gives:
\begin{equation}
\mathbf{\bar{v}}_{g r} = \frac{1}{N} \sum_{i=1}^N \mathbf{\bar{v}}_{g r}^{i}, 
\label{eq.N2}
\end{equation} 
where $\mathbf{\bar{v}}_{g r}^{i}$ means the $i$-th product in the $gr$ list. Similarly, the mean textual representation of the $gr$ list is denoted as $\mathbf{\bar{w}}_{g r}$, which is calculated as $\mathbf{\bar{v}}_{g r}$ in eq. (\ref{eq.N2}).
The dot product is applied to calculate the content-based similarity in terms of each visual and textual modality. The relative importance of the two modalities is controlled by the same trade-off parameters $\pi$ and $1-\pi$, which leads to the following product matching consistency regularization:

\begin{equation}
\begin{aligned}
s_{g r}^c=\pi\left(\mathbf{\bar{v}}_{g r}\right)^T \mathbf{\bar{v}}_r^{c_g} +(1-\pi)\left(\mathbf{\bar{w}}_{g r}\right)^T \mathbf{\bar{w}}_r^{c_g}.
\label{eq.stc}
\end{aligned}
\end{equation}
$s_{g r}^c$ is the product matching consistency score and the subscript $c$ differentiates it from the product matching score $s_{g r}$ in eq. (\ref{eq.stb}), which the product matching or product matching relationship is derived from all product-pair interactions. In eq. (\ref{eq.stc}), a higher similarity value of $s_{g r}^c$ indicates stronger regularization of product-matching consistency when recommending a suitable matching product. The product matching consistency score is used as auxiliary information to improve prediction outcomes, especially when the candidate matching product $r$ has little historical product-product interactions.

\subsection{Optimization and Overall Preference Prediction}
The above four components, including $s_{u r}$ -- the user personal preference score, $s_{g r}$ -- the product matching score, and $s_{u r}^c$ and $s_{g r}^c$ -- the two consistency regularization scores, are linearly combined, illustrated in Fig. 2, as the overall CR-BPR model. In other words, to predict the overall personalized product matching preference, eq. (\ref{eq:overall_obj}) is formulated as follows:
\begin{equation}
p^{u,g}_r=\mu \cdot s_{g r}+(1-\mu) \cdot s_{u r}+\phi \cdot s_{g c}^c+\varphi \cdot s_{u r}^c
\label{eq.overall}
\end{equation}
$p_r^{u,g}$ shows the preference score of user $u$ towards product $r$ to match with a given garment $g$. 
In eq. (\ref{eq.overall}), the trade-off parameter $\mu$ is applied to balance the two main components of user preference and product matching, while $\phi$ and $\varphi$ control the contributions of the two consistency regularization components. 

\textbf{Loss function: }
The CR-BPR model (eq. (\ref{eq.overall})) is optimized with the BPR loss, which is a pairwise loss to enlarge the gap between the positive and negative samples, thereby ranking the positive candidates higher. Let the preference scores for a pair of positive and negative samples, be obtained using eq. (\ref{eq.overall}), are denoted as $p^{u,g}_{r_+}$ and $p^{u,g}_{r_-}$, respectively. The BPR loss on the whole training set $\mathcal{D}$ is calculated by:
\begin{equation}
\begin{aligned}
\mathcal{L}=\sum_{\mathcal{D}}\left[-\ln \left(\sigma\left(p_{r_+}^{u,g}-p_{r_-}^{u,g}\right)\right)\right]+\frac{\lambda}{2}\left\|\Theta_F\right\|_F^2,
\end{aligned}
\end{equation}
where $\mathcal{D} = \left\{(u, g, r_+, r_-) \mid u \in \mathcal{U} \wedge g \in \mathcal{G} \wedge (r_+, r_-) \in \mathcal{R} \right\}.$ 
${\lambda}$ is the non-negative hyperparameter and ${\Theta_F}$ represents the set of parameters of the model. 

 \textbf{Feature scaling based parameter update. }
 It is important to note that in other studies, such as~\cite{GPBPR}, the representation of target matching product $r$ in both user preference modeling (eq. (\ref{eq.sub})) and product matching modeling (eq. (\ref{eq.stb})) are assumed the same, namely $\mathbf{v}_r^p=\mathbf{v}_r^m$ and $\mathbf{w}_r^p=\mathbf{w}_r^m$. Comparatively, they are different in current formulation of CR-BPR in order to improve matching product representation in multiple latent spaces and capture relations. The representation $\mathbf{v}_g$ is adjusted in the product matching branch and $\mathbf{v}_u$ in the user preference branch as follows.
\begin{subequations}
\label{ori_update}
\begin{eqnarray}
\mathbf{v_g} \leftarrow \mathbf{v_g}+\rho \cdot\left(\sigma\left(-p_{r_+ r_-}^{u,g}\right) \cdot\left(\mathbf{v_{r_+}^m}-\mathbf{v_{r_-}^m}\right)-\lambda \mathbf{v_g}\right) \\
\mathbf{v_u} \leftarrow \mathbf{v_u}+\rho \cdot\left(\sigma\left(-p_{r_+ r_-}^{u,g}\right) \cdot\left(\mathbf{v_{r_+}^p}-\mathbf{v_{r_-}^p}\right)-\lambda \mathbf{v_u}\right)
\end{eqnarray}
\end{subequations}
This feature scaling based parameter update (eq. (\ref{ori_update})) can improve model learning. In later section \ref{sec:FS_analysis}, this will compare with baseline model GP-BPR~\cite{GPBPR}, in which $\mathbf{v}_g$ and $\mathbf{v}_u$ are updated without any feature scaling. 

\section{Experiments and Result Discussion}\label{sec_4}
To evaluate the proposed method, detailed experiments and analyses involving different datasets were carried out. 


\subsection{Experiments}
\subsubsection{Datasets}
The proposed model is evaluated on two benchmark fashion datasets composed of real-world interaction data: IQON3000~\cite{GPBPR} and Polyvore-519~\cite{binary}. The exact same version of IQON3000 of~\cite{GPBPR} was used in the current study, including data distribution and content features. For Polyvore-519, triplet data samples of user and fashion product pairs were created from user-outfit interaction data, keeping only product pairs of one top garment and one bottom garment in each outfit and discarding all other product categories. Since Polyvore-519 dataset has limited interaction data, it was further refined by removing those user-product pair samples with less than two interactions. 
Detailed statistics of the two datasets are given in Table \ref{dataset}.

\begin{table}[!t]
    \caption{Statistics of datasets being used as a basis for the experimental analysis and validation.}
    \centering
    \setlength{\tabcolsep}{5mm}
    \begin{tabular}{lcc}
        \toprule
         Dataset &Polyvore-519 & IQON3000  \\
        \hline
        Number of users & 519 & 3236  \\
        Number of top garments & 16887 & 99655   \\
        Number of bottom garments & 16220 & 43086  \\
        Total number of garments & 33107 & 142737  \\
        \hline      
        Number of training samples & 39133 & 170601 \\
        Number of validation samples & 5110 & 23095  \\
        Number of testing samples & 5197 & 23095  \\
        Total number of samples & 49440 & 216791 \\
        \hline       
    \end{tabular}
    \label{dataset}
\end{table}

The proposed CR-BPR recommends matching clothing by learning multi-modal representations of clothing products, which, in particular, cover visual and textual features. The available features provided by~\cite{GPBPR} for IQON3000 are applied, these being 2048-D visual features extracted from the pre-trained ResNet50 and 400-D textural features extracted from a TextCNN.
For the Polyvore-519, the 2400-D textural features are available as in the original dataset, which were extracted using a pre-trained \textup{AlexNet}~\cite{alexnet}. For visual features, the Resnet152~\cite{resnet50} pre-trained on ImageNet was used to obtain a 2048-D visual feature for each clothing product. 

\subsubsection{Implementation details and evaluation metrics}
The adaptive moment estimation method (Adam) optimizer~\cite{ADAM} was used to train the CR-BPR model, with the maximum training epoch being set to 80, but an early stopping strategy was applied with the patient parameter being set to 8 epochs. Each triplet data from the training set is a positive sample. 
For model training purpose, another bottom product will be randomly selected from the corresponding product set 
as a negative sample. The experiment was conducted first to recommend a target matching bottom product $r$, then for recommending a target top product to match with a specified bottom.

The grid search strategy was applied to search for the best settings for all the training parameters. More specifically, the mini-batch size is searched in the range of [64, 128, 256, 512], weight decay in the range of [0.001, 0.0001, 0.00001, 0.000001, 0.0000001], hidden dimension in the range of [256, 512], and learning rate in the range of [0.01, 0.001, 0.0001], respectively. The layer ($K$) of the multi-purpose projection was set to 1, while the hyperparameters of the model were set differently for the two datasets based on empirical observations.
Following~\cite{GPBPR}, the visual and textual modalities are assumed equally important, thus the trade-off parameters $\eta$ and $\pi$ were set to 0.5 in all experiments.
The consistency modeling was based on the interaction set of users and products in the training set. 
We have included 4 different \textbf{evaluation metrics} to fully illustrate the effectiveness of our proposed CR-BPR: AUC, HR@K, NDCG@K, and MRR@K. 
The Area Under the Curve of ROC (AUC)~\cite{AUC} was adopted as the metric to evaluate the performance of different models in terms of matching garment prediction. The AUC measures the proportion on the whole testing/validation set that the model successfully selects the positive matching garment over the negative one. The best performance of the model is determined to be the AUC result on the testing set at the epoch when the validation set achieves its best performance.
Hits Ratio, HR@K, assesses the proportion of relevant products that rank among the top recommendations,
indicating user satisfaction. 
Furthermore, Normalized Discounted Cumulative Gain, NDCG@K, is a ranking metric that takes into account both the relevance of the products and their positions in the lists, while Mean Reciprocal Rank, MRR@K, provides a straightforward measure on the ranking performance by only considering the position of the first relevant product in the list. In our evaluation, we chose K=10 for reporting results, facilitating comparison with other literature. 
By including these metrics, we show how the CR-BPR model is capable of capturing user preferences and provide highly accurate recommendations, which allows for a more comprehensive evaluation of the model’s effectiveness.


\subsection{Performance Comparison by Quantitative Evaluations}
To demonstrate the effectiveness of the proposed CR-BPR, it was compared with the existing methods of personalized clothing matching including the following:
\begin{itemize}
\item \textbf{MF-BPR}~\cite{BPR}, which applies the Matrix Factorization (MF) to capture the latent user-product relations with the Bayesian Personalized Ranking (BPR) algorithm.
\item \textbf{V-BPR}~\cite{VBPR}, is an extension of MF-BPR, which integrates visual features into the BPR model and captures specific visual preferences of users. 
\item \textbf{T-BPR}~\cite{GPBPR}, which adopts the same algorithm as V-BPR, replacing the visual features with textural features to leverage the textual information into the BPR modeling. 
\item \textbf{VT-BPR}~\cite{GPBPR}, which combines V-BPR and T-BPR by further comprehensively characterizing user preferences on the basis of both visual and textual factors.
\item \textbf{GP-BPR}~\cite{GPBPR}, which is a multi-modal feature-based personalized compatibility modeling method. It effectively models both user preference and product matching relations with a joint BPR framework form. This is the baseline of the present CR-BPR model. 
\item \textbf{PCE-Net} \cite{PCE} adopts attention-based product matching embedding modeling and attention-based personal preference modeling to further enhance performance.
\item \textbf{CP-TransMatch} \cite{CP} uses a single-component translation operation and two graph learning modules to effectively model higher order user-product-product interactions within a multi-relational graph. In Table~\ref{table_overall} we abbreviate it as \textbf{TransMatch}.
\end{itemize}

\begin{table*}[t]
\caption{Performance comparison. \textbf{Bold} data indicate the best results, and \underline{\textit{underlined}} ones are the second best results.}
\begin{center}
\setlength{\tabcolsep}{0.7mm}
\begin{tabular}{{ll|l|cccccccc|cc}}
\toprule
\multicolumn{2}{c|}{Dataset} &Metrics  & MF-BPR & T-BPR & V-BPR &VT-BPR  & GP-BPR & TransMatch & PCE-NET &\textbf{CR-BPR} & \#Improve \\
\cmidrule(r){1-3} \cmidrule[\arrayrulewidth](lr){4-10} \cmidrule(lr){11-12}
\multirow{8}{*}{\rotatebox{90}{With Specified Top}} & \multirow{4}{*}{\rotatebox{90}{Polyvore}}
& AUC ($\uparrow$)& 0.7639  &0.7839 &0.8074 &0.8105 &0.8232  &\underline{\textit{0.9001}} &0.8235 &\textbf{0.9708} & 7.85\%  \\
~&~&HR@10 ($\uparrow$)&0.6916 &0.6502 &0.7141 &0.6846 &0.7121  &\underline{\textit{0.8573}} &0.4208 & \textbf{0.9582}&11.77\%\\
~&~&NDCG@10 ($\uparrow$)&0.5434 &0.4868 &\underline{\textit{0.5848}} &0.5266 &0.5716  &0.5472 &0.2380 &\textbf{0.8717} & 49.05\%\\
~&~&MRR@10 ($\uparrow$)&0.4973 &0.4359 &\underline{\textit{0.5443}} &0.4772 &0.5277  &0.5144 & 0.1825 &\textbf{0.8361} &53.61\% \\
\cmidrule(r){2-3} \cmidrule(lr){4-10} \cmidrule(lr){11-12}
~&\multirow{4}{*}{\rotatebox{90}{IQON}}
&AUC ($\uparrow$)&0.8309 &0.8316 & 0.8356 &0.8384 &0.8569  &\underline{\textit{0.8842}} &0.8341 &\textbf{0.9737} &10.12\% \\
~&~&HR@10 ($\uparrow$)&0.7024 &0.6361 &0.6941 &0.7003 &0.7396  &\underline{\textit{0.8789}}  &0.6399 &\textbf{0.9505} &8.15\% \\
~&~&NDCG@10 ($\uparrow$)&0.5243 &0.4301 &0.5230 &0.5134 &0.5849  &\underline{\textit{0.6453}} & 0.4110&\textbf{0.8746} &35.53\% \\
~&~&MRR@10 ($\uparrow$)&0.4687 &0.3666 &0.4694 &0.4551 &0.5366  &\underline{\textit{0.5430}} & 0.3399&\textbf{0.8518} &56.86\% \\
\midrule
\multirow{8}{*}{\rotatebox{90}{With Specified Bottom}} & \multirow{4}{*}{\rotatebox{90}{Polyvore}}
& AUC ($\uparrow$)& 0.5724 & 0.6250 &0.6644 &0.6785  &0.7075  &\underline{\textit{0.7281}} &0.7072 &\textbf{0.7637} &4.89\%  \\
~&~&HR@10 ($\uparrow$)&0.2598 &\underline{\textit{0.2938}} &0.2576  &0.2523 & 0.2500 &0.2859 &0.2222 &\textbf{0.3689} &25.56\% \\
~&~&NDCG@10 ($\uparrow$)&0.1331 &0.1638 &0.1316 &0.1263 &0.1253  &\underline{\textit{0.1993}} &0.1120 &\textbf{0.2026} &1.71\% \\
~&~&MRR@10 ($\uparrow$)&0.0950 &0.1243 &0.0937 &0.0885 &0.0877 &\underline{\textit{0.1419}} &0.0789 &\textbf{0.1522} &7.26\% \\
\cmidrule(r){2-3} \cmidrule(lr){4-10} \cmidrule(lr){11-12}
~&\multirow{4}{*}{\rotatebox{90}{IQON}}
&AUC ($\uparrow$)&0.6849 &0.7510&0.7626 &0.7890  &0.7913 &\underline{\textit{0.8523}}  &0.7585 &\textbf{0.9031} & 5.96\% \\
~&~&HR@10 ($\uparrow$)&0.5598  &0.5315& 0.5609&0.5341  &0.6137  &\underline{\textit{0.6251}} &0.5111 &\textbf{0.8357} & 33.69\%\\
~&~&NDCG@10 ($\uparrow$)&0.3776 &0.3493 &0.3724 &0.3576  & \underline{\textit{0.4663}} &0.3756 &0.3239 &\textbf{0.7863} &68.63\% \\
~&~&MRR@10 ($\uparrow$)&0.3209 &0.2929 &0.3141 &0.3032 & \underline{\textit{0.4196}} &0.2989 &0.2669 &\textbf{0.7709} &83.72\% \\
\hline
\end{tabular}\\
\end{center}
\label{table_overall}
\end{table*}

The above methods are chosen for comparison because they are most relevant to our specific task, without requiring additional information, such as product attributes or user profiles, as that in~\cite{ma2024personalized}. Among the selected methods, \textbf{TransMatch}~\cite{CP} presents the state-of-the-art approach, surpassing more recent models \cite{liu2024unifying, liu2024outfit}. 
The overall performance of all compared methods on two datasets
are listed in Table \ref{table_overall}, where \textit{Specified Top} represents the given product is a top garment, and the matching product to be recommended is a bottom garment, and vise versa.
We have the following observations from Table~\ref{table_overall}:

\begin{itemize}
    \item Our proposed CR-BPR achieves the best performance across all datasets under two specified settings for all four evaluation metrics, extensively demonstrating its effectiveness.
    \item MF-BPR, T-BPR, V-BPR, and VT-BPR perform the worst among all methods, as they only model user-product interaction patterns while ignoring product-product compatibility patterns. In comparison, GP-BPR, incorporating both user preference modeling and product matching modeling,  outperforms the MF-BPR, T-BPR, V-BPR, and VT-BPR, which illustrate that both two relation modeling contribute significantly in the task of personalized complementary clothing recommendation.
    \item From the comparison between T-BPR and V-BPR, we can see that V-BPR performs better than T-BPR in the most scenarios, which indicates that visual information contributes more significantly to this task.  
    Furthermore, the models that used both visual and textual features (i.e. VT-BPR, GP-BPR, and PCE-NET) achieved better performance than models that only used single modality, e.g. V-BPR and T-BPR. The results showed that textual product descriptions and visual images are both critical in exploring user preferences and clothing matching. 
    \item In addition, it is evident that the GP-BPR and PCE-NET came cloest to the present CR-BPR method, confirming the benefits of modeling both user preference with user-product interactions in an integrated framework.
    \item TransMatch models multi-relational connectivity between fashion products subject to users by a multi-edge graph, and performs the second best in the most settings compared to other baselines. This illustrates that fully exploring the user connections through user-product-product third order transactions benefits user preference modeling thoroughly. However, the complex path-based information translation, although with increased computation difficulty, can easily lead to overfitting and reduced effectiveness, especially when the dataset is imbalanced. 
\end{itemize}


\subsection{Ablation Study} 
\label{sec: ablation}

With reference to Fig.~\ref{fig:method}, CR-BPR integrates four branches: two main branches for user preference and product matching, and two consistency-regularization branches for user-preference and product-matching consistency. Compared to the baseline GP-BPR model, the present CR-BRP model introduces feature scaling procedure (w.r.t. eqs. (\ref{eq:output_feature}) and (\ref{eq:norm})) and two consistency regularization modules (w.r.t. eqs. (\ref{eq.suc}) and (\ref{eq.stc})). Comprehensive ablation study was conducted to validate the effectiveness of feature scaling and consistency regularizations in the following subsections.

\subsubsection{\textbf{Effectiveness on Feature Scaling}}\label{sec:FS_analysis}


\begin{table*}[t]
\caption{Modality and Feature Scaling Ablation Experiment. -w/o refers to the evaluation results of the models WITHOUT applying relevant information/features.}
\begin{center}
\setlength{\tabcolsep}{3mm}
\begin{tabular}{{ll|l|ccccc||c}}
\toprule[\arrayrulewidth]
\multicolumn{2}{c|}{Dataset} & Metrics & -w/o FS* & -w/o V & -w/o T & -w/o V+FS* & -w/o T+FS* & \textbf{CR-BPR}  \\
\cmidrule(r){1-3} 
\cmidrule[\arrayrulewidth](lr){4-8} 
\cmidrule(lr){9-9}

\multirow{8}{*}{\rotatebox{90}{With Specified Top}} & \multirow{4}{*}{\rotatebox{90}{Polyvore}}
& AUC ($\uparrow$)&0.8157 &0.8894 &0.9683 &0.7622 &0.8099 &\textbf{0.9708} \\

~&~&HR@10 ($\uparrow$)&0.3933 &0.5915 &0.8686 &0.3063 &0.4172 &\textbf{0.9582} \\

~&~&NDCG@10 ($\uparrow$)&0.2165 &0.3490 &0.6233 &0.1634 &0.2337 &\textbf{0.8717} \\

~&~&MRR@10 ($\uparrow$)&0.1628 &0.2736 &0.5451 &0.1198 &0.1779 &\textbf{0.8361} \\

\cmidrule{2-3} \cmidrule[\arrayrulewidth](lr){4-8} \cmidrule(lr){9-9}

~&\multirow{4}{*}{\rotatebox{90}{IQON}}
& AUC ($\uparrow$)&0.8087 &0.9391 &0.9472 &0.7692 &0.8176 &\textbf{0.9737} \\

~&~&HR@10 ($\uparrow$)&0.5791 &0.8879 &0.9034 &0.5053 &0.6087 &\textbf{0.9505} \\

~&~&NDCG@10 ($\uparrow$)&0.3395 &0.7161 &0.7808 &0.2909 &0.3708 &\textbf{0.8746} \\

~&~&MRR@10 ($\uparrow$)&0.2662 &0.6608 &0.7416 &0.2256 &0.2975 &\textbf{0.8518} \\

\cmidrule(lr){1-3} \cmidrule[\arrayrulewidth](lr){4-8} \cmidrule(lr){9-9}

\multirow{8}{*}{\rotatebox{90}{With Specified Bottom}} & \multirow{4}{*}{\rotatebox{90}{Polyvore}}
& AUC ($\uparrow$) &0.6990 &0.7447 &0.7597 &0.6117 &0.6966 &\textbf{0.7637} \\

~&~&HR@10 ($\uparrow$)&0.2609 &0.3325 &0.3679 &0.1965 &0.2500 &\textbf{0.3689} \\

~&~&NDCG@10 ($\uparrow$)&0.1402 &0.1747 &0.2065 &0.1052 &0.1301 &\textbf{0.2026} \\

~&~&MRR@10 ($\uparrow$)&0.1038 &0.1269 &0.1571 &0.0777 &0.0940 &\textbf{0.1522} \\

\cmidrule{2-3} \cmidrule[\arrayrulewidth](lr){4-8} \cmidrule(lr){9-9}

~&\multirow{4}{*}{\rotatebox{90}{IQON}}
& AUC ($\uparrow$)&0.8120 &0.7990 &0.8504 &0.6883 &0.7451 &\textbf{0.9031} \\

~&~&HR@10 ($\uparrow$)&0.6809 &0.6491 &0.7051 &0.3861 &0.4713 &\textbf{0.8357} \\

~&~&NDCG@10 ($\uparrow$)&0.5122 &0.4699 &0.4914 &0.2153 &0.2822 &\textbf{0.7863} \\

~&~&MRR@10 ($\uparrow$)&0.4594 &0.4139 &0.4250 &0.1635 &0.2246 &\textbf{0.7709} \\
\hline
\end{tabular}
\end{center}
\label{tab:ablation_part1}
\end{table*}

Defined in eqs. (4a) and (4b) in the paper, the feature scaling we adopted is to address the issue of significant scale imbalance caused likely by different data sources and different feature spaces. For example, the large scale of color vectors in clothing visual embeddings, may cause the model to overly focus on color when computing similarity with latent embeddings of matching products, while neglects other important features. To mitigate this bias, we apply feature scaling, encouraging the model to focus on the relevant features in a more balanced manner. 

\textit{Comparison of different modalities with and without feature scaling in our CR-BPR.} To explore the effect of feature scaling on our proposed CR-BPR, we compares in Table~\ref{tab:ablation_part1} that our CR-BPR with and without Visual/Textual information, and with and without feature scaling on each kind of modality. Specifically, -w/o FS* means our CR-BPR without Feature Scaling on all four branches.
-w/o V and -w/o T represent without exploring Visual and Textual modality of fashion products, respectively. In addition, -w/o V+FS* and -w/o T+FS* denote without Visual/Textual modality and Feature Scaling on all branches. From the experiments we can conclude that Feature Scaling contributes very significantly in our CR-BPR, even when with only one modality. Furthermore,
Based on the comparison of experiments - w/o V and - w/o T, we observe that the evaluation results significantly decline when the visual modality is excluded from the model. In contrast, the ablation experiment with the textual modality shows that the test results do not decline as much. Therefore, we can conclude that in our model, visual features provide more effective and valuable matching information compared to textual features.

\begin{table}[th]
\caption{Performance comparison in terms of prediction AUC ($\uparrow$) between the original baselines and those with applied Feature Scaling (+FS*).}
\begin{center}
\setlength{\tabcolsep}{5mm}{
\begin{tabular}{{l|cc|cc}}
\toprule
\multirow{2}{*}{Methods}  & \multicolumn{2}{c|}{With Specified Top} &\multicolumn{2}{c}{With Specified Bottom}\\
~ & Polyvore & IQON3000 & Polyvore & IQON3000 \\
\midrule
MF-BPR~\cite{BPR} & 0.7639 & 0.8309 & 0.5724 & 0.6849\\
T-BPR~\cite{GPBPR}  & 0.7839 & 0.8316 & 0.6250 & 0.7510\\
V-BPR~\cite{VBPR} & 0.8074 & 0.8356 & 0.6644 & 0.7626\\
VT-BPR~\cite{GPBPR}  & 0.8105 & 0.8384 & 0.6785 & 0.7890\\
GP-BPR~\cite{GPBPR} & 0.8232 & 0.8569 & 0.7075 & 0.7913\\
\midrule
MF-BPR + FS* & 0.7375 & 0.8155 & 0.5957 & 0.6652 \\
T-BPR + FS* & 0.7273 & 0.9463 & 0.6290 & 0.8653 \\
V-BPR + FS* & 0.7443 & 0.8656 & 0.6523 & 0.7312 \\
VT-BPR + FS* & 0.7549 & 0.9347 & 0.6454 & 0.8528 \\
GP-BPR + FS* & 0.8863 & 0.9528  & 0.7341 & 0.8899\\
\hline
\end{tabular}}\\
\end{center}
\label{tab：FS}
\end{table}

\begin{figure}[th]
  \centering
   \includegraphics[width=\linewidth]{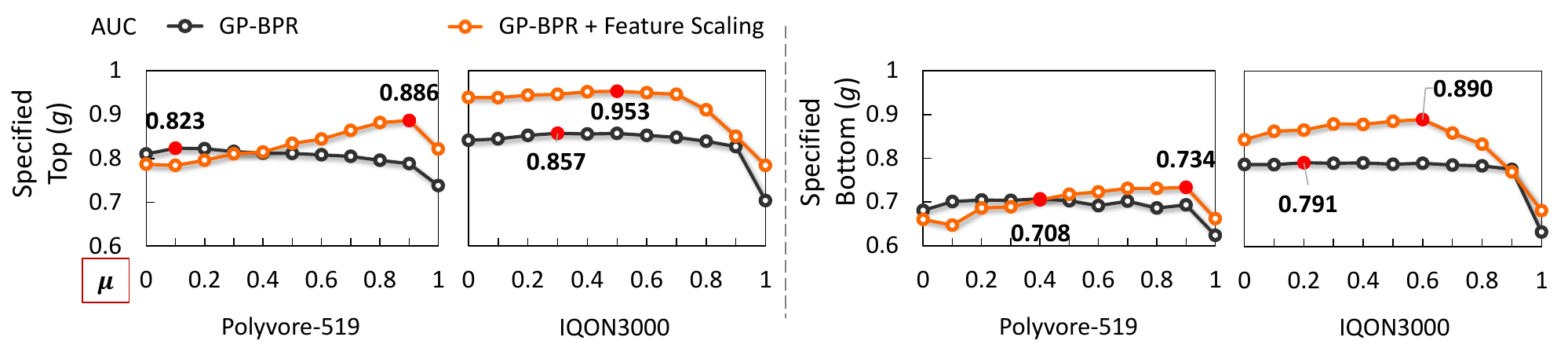}
  \caption{The performance comparison between the GP-BPR baseline and GP-BPR with Feature Scaling using the Polyvore-519 and IQON3000 datasets when (a) specify top and recommend bottom, and (b) specify bottom and recommend top. $\mu$ is the weight of Product Matching modeling. }
  \label{gpbpr.norm}
\end{figure}

\textit{Comparison of different baselines with and without feature scaling.} As compared in Table~\ref{tab：FS}, feature scaling improves  most baseline models with multi source data, such as GP-BPR, especially in IQON3000 dataset, whereas the benefit is not so obvious for the Polyvore-519 dataset.
The magnitude, range, and units of features always vary in different datasets. Even though feature scaling has been proven to be advantageous in many machine learning applications, its application in practice will depend upon the specific problem being addressed. For example, models which only involve one specific type of data (such as MF-BPR) may not benefit from feature scaling whereas those which involve multiple types of data may benefit. 
GP-BPR distinguishes itself from other methods in that it utilizes a diverse range of data. Through experiments on both datasets in cases of top and bottom products as the specified products, it was observed that feature scaling provided a significant benefit to GP-BPR.

To further investigate the detailed impacts on \textbf{our baseline} GP-BPR model, the AUC of GP-BPR with and without \textit{Feature Scaling} was assessed in Fig.~\ref{gpbpr.norm}, against different settings of trade-off parameter $\mu$ of eq. (\ref{eq.overall}). GP-BPR only considers product matching (i.e. product-matching score) $s_{g r}$ and user preference (i.e. user preference score) $s_{u r}$. The horizontal axis $\mu$ of Fig.~\ref{gpbpr.norm} indicates the weight of the product-matching modeling, according to eq. (\ref{eq.overall}). 
It can be seen from Fig.~\ref{gpbpr.norm} that user preference dominates the overall performance in the original GP-BPR. After feature scaling, however, the performance of all experiments improved significantly, with the product-matching branch contributing more to the overall model performance. Especially in IQON3000 dataset, both branches received balanced contributions. 
Furthermore, when $\mu$ equals 1, there was a significant drop in performance. It implies that general product-matching modeling performs poorly in personalized clothing matching modeling, and both product matching and user preference modeling branches have significant contributions to the performance.

\begin{table*}[t]
\caption{Ablation Experiment. -w/o refers to the evaluation results of the models WITHOUT applying relevant modules.}
\begin{center}
\setlength{\tabcolsep}{4mm}
\begin{tabular}{{ll|l|cccc||c}}
\toprule[\arrayrulewidth]
\multicolumn{2}{c|}{Dataset} & Metrics & -w/o UC & -w/o GC & -w/o UC+U & -w/o GC+G & \textbf{CR-BPR}  \\
\cmidrule(r){1-3} 
\cmidrule[\arrayrulewidth](lr){4-7} 
\cmidrule(lr){8-8}

\multirow{8}{*}{\rotatebox{90}{With Specified Top}} & \multirow{4}{*}{\rotatebox{90}{Polyvore}}
& AUC ($\uparrow$)&0.9609 &0.9609 &0.8157 &0.9542 &\textbf{0.9708} \\

~&~&HR@10 ($\uparrow$)&0.9350 &0.9173 &0.3933 &0.8676 &\textbf{0.9582} \\

~&~&NDCG@10 ($\uparrow$)&0.7174 &0.7363 &0.2165 &0.5019 &\textbf{0.8717} \\

~&~&MRR@10 ($\uparrow$)&0.6467 &0.6772 &0.1628 &0.3868 &\textbf{0.8361} \\

\cmidrule{2-3} \cmidrule[\arrayrulewidth](lr){4-7} \cmidrule(lr){8-8}

~&\multirow{4}{*}{\rotatebox{90}{IQON}}
& AUC ($\uparrow$)&\textbf{0.9738} &0.9727 &0.8087 &0.9040 &0.9737 \\

~&~&HR@10 ($\uparrow$)&0.9092 &0.9108 &0.5791 &0.8351 &\textbf{0.9505} \\

~&~&NDCG@10 ($\uparrow$)&0.7655 &0.7752 &0.3395 &0.6519 &\textbf{0.8746} \\

~&~&MRR@10 ($\uparrow$)&0.7194 &0.7314 &0.2662 &0.5929 &\textbf{0.8518} \\

\cmidrule(lr){1-3} \cmidrule[\arrayrulewidth](lr){4-7} \cmidrule(lr){8-8}

\multirow{8}{*}{\rotatebox{90}{With Specified Bottom}} & \multirow{4}{*}{\rotatebox{90}{Polyvore}}
& AUC ($\uparrow$)&0.7606 &0.7554 &0.6990 &0.6931 &\textbf{0.7637} \\

~&~&HR@10 ($\uparrow$)&0.3379 &0.3198 &0.2609 &0.2682 &\textbf{0.3689} \\

~&~&NDCG@10 ($\uparrow$)&0.1841 &0.1672 &0.1402 &0.1427 &\textbf{0.2026} \\

~&~&MRR@10 ($\uparrow$)&0.1374 &0.1212 &0.1038 &0.1047 &\textbf{0.1522} \\

\cmidrule{2-3} \cmidrule[\arrayrulewidth](lr){4-7} \cmidrule(lr){8-8}

~&\multirow{4}{*}{\rotatebox{90}{IQON}}
& AUC ($\uparrow$)&0.9018 &0.8993 &0.8120 &0.7225 &\textbf{0.9031} \\

~&~&HR@10 ($\uparrow$)&0.5667 &0.6219 &0.6809 &0.5099 &\textbf{0.8357} \\

~&~&NDCG@10 ($\uparrow$)&0.3928 &0.4053 &0.5122 &0.3047 &\textbf{0.7863} \\

~&~&MRR@10 ($\uparrow$)&0.3387 &0.3376 &0.4594 &0.2412 &\textbf{0.7709} \\
\hline
\end{tabular}
\end{center}
\label{tab:ablation_part2}
\end{table*}

\subsubsection{\textbf{Effectiveness of Consistency Regularization}}
\label{sec:CR_analysis}

Table \ref{tab:ablation_part2} shows the overall ablation experiment results.
The CR-BRP model has two consistency regularization modules, we compare the model performance with only one consistency regularization module, namely CR-BPR without UC and denotes as `CR-BPR w/o UC', and without GC denoted as `CR-BPR w/o GC'. 
In addition to CR-BPR w/o UC and w/o GC, other supplementary results further elaborate the model performance under different settings. Specifically, -w/o UC+U refers to the evaluation results of the CR-BPR model WITHOUT applying both UC branch and User preference modeling (U) branch. On the other hand, -w/o GC+G refers to the results of the CR-BPR WITHOUT applying both GC and product matching modeling (G) branch. 

Ablation experiment results in Table \ref{tab:ablation_part2} clearly show that the removal of any single module will result in a decline in performance across all metrics. This highlights the significance of each component in contributing to the model's effectiveness. The best results are achieved when all four modules are integrated, indicating that the combined approach yields the most accurate recommendations.
We also find that -w/o GC+G performs better than -w/o UC+U in almost all context, in other words, which in contrast indicates CR-BPR without exploring the whole Product Matching and Product Matching Consistency modeling contribute more for the overall recommendation performance.
In addition, our CR-BPR model is more robust in case of rich data information, such as IQON3000. From the comparison between -w/o UC and -w/o UC+U, and the the comparison between -w/o GC and -w/o GC+G, it has generally validated the benefits of consistency regularization, in terms of both user-preference and product-matching consistency.

\subsection{Qualitative Evaluations}
\begin{figure}[htb]
    \centering
    \includegraphics[width=\linewidth]{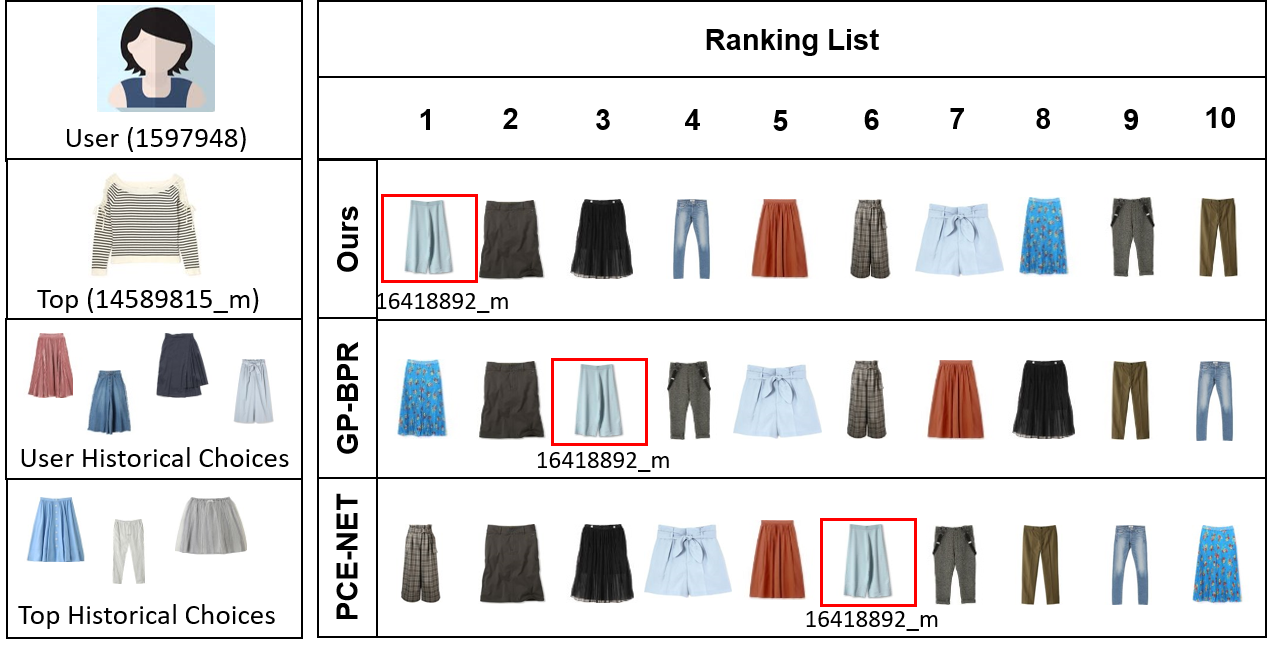}
    \caption{Illustration of the clothing matching recommendation results provided by three methods. The ground-truth are highlighted by red frame.}
    \label{fig:case_study}
\end{figure}

\begin{figure}[htb]
    \centering
    \includegraphics[width=\linewidth]{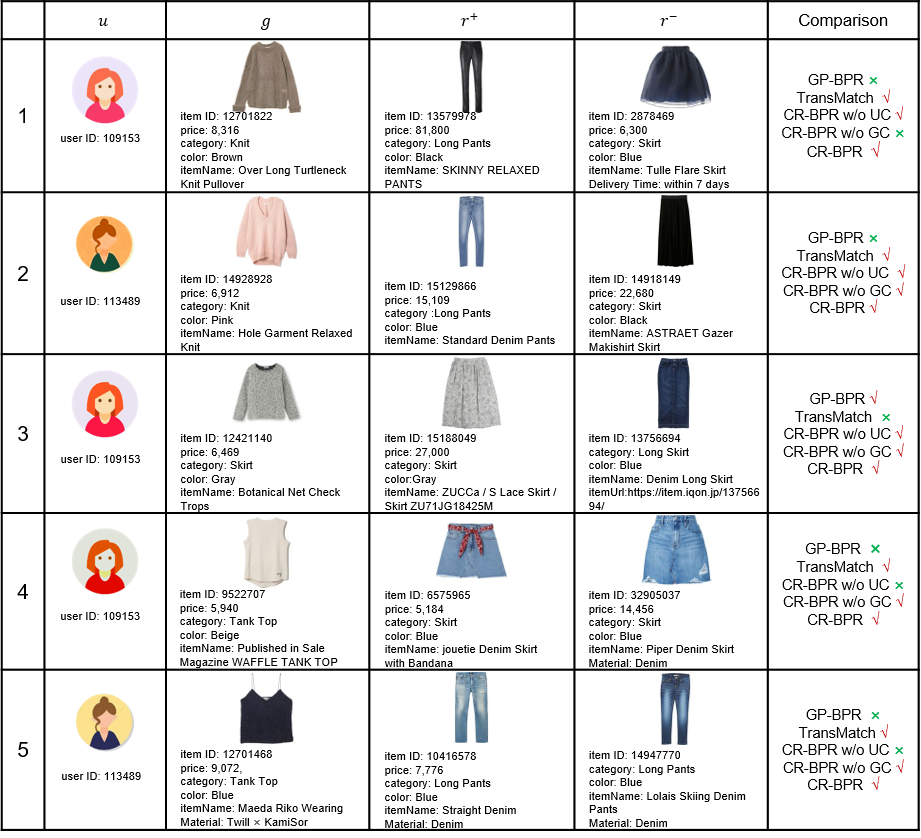}
    \caption{Comparison of the clothing matching recommendation results provided by different methods. }
    \label{fig:case_study_2}
\end{figure}

Although user study, involving human judgment, has been used in evaluating recommendation algorithms~\cite{PCE}, user study in not applicable or being a suitable strategy in the current study for personalized complementary-product recommendations for two reasons.
First, user study is more illustrative rather than essential in most cases as users may be subjective and biased, meanwhile the sample size of user evaluation is often limited, making user evaluation lack of comprehensiveness. 
Second, user study evaluation can only focus on product matching instead of user preference because the latter needs historical transaction/browsing data.
In this regard, to verify the practical application value of our CR-BPR, we conducted qualitative evaluations by retrieving matching products for comparing and visualizing some test examples, following the evaluation strategy widely adopted by previous similar studies~\cite{GPBPR, PAI, CP,liu2024unifying, liu2024outfit}.

Fig. \ref{fig:case_study} illustrates the clothing matching ranking list results generated from three different methods: our CR-BPR, GP-BPR, and PCE-NET. A user and top garment pair in testing dataset was randomly selected from IQON3000 as query, and 9 bottom garments were randomly selected combining with the ground-truth matching product as the candidates, in which the ground-truth was highlighted in red frame. By combining user and top historical choices, the ranking results show that our method outperforms the other two methods not only by ranking ground-truth products higher but also by recommending products that are more similar to the target product. 

Another comparison results in respect of recommendation accuracy are illustrated in Fig. \ref{fig:case_study_2}. Specifically, we randomly selected several user-product-product transaction triplets, denoted as $<u, g, r_+>$, from the testing dataset. In each triplet, we chose a negative matching product ($r_-$) by selecting a product that the user had not interacted with before. The aim is to correctly match the right bottom clothing for each user with the given top garment.
Each top and bottom clothing are represented by visual images and textual descriptions, while user identities are anonymized to protect privacy, with only ID information utilized. As can be seen from Fig.~\ref{fig:case_study_2} that our CR-BPR performs well across all cases. 
In addition, we observed that matching predictions are easier when the given tops and bottoms have some visual or textual connection. This implies that the presence of distinct and recognizable features in both visual and textual representations facilitates the matching process. However, when two matching products are very similar in their semantic features, it becomes challenging for traditional compatibility models, such as GP-BPR and CR-BPR without user connections (i.e., CR-BPR w/o UC), to distinguish between the two products. This difficulty arises because these traditional models rely heavily on the inherent features of the products themselves, without considering the additional context that user interactions can provide.
By further incorporating user implicit connections, such as TransMatch and our CR-BPR, it help the model to make more accurate predictions beyond the general compatibility modeling.
The results further demonstrate the effectiveness of our proposed method in addressing personalized complementary clothing recommendation task.

\subsection{Discussions}
In the two consistency regularization branches, the weight $\phi$ of GC branch, the weight $\varphi$ of UC branch (eq. (\ref{eq.overall})), and the number of historical choices ($N$) (eqs. (\ref{eq.N}) and (\ref{eq.N2})) are the most important factors to influence the final performance. 
In the context of our study, we performed further experiments fine-tuning the above parameters, thereby maximizing our CR-BPR effectiveness in personalized clothing matching.

\subsubsection{\textbf{Balancing Weights of the Two Consistency Regularizers ($\phi,\varphi$).}}

\begin{figure}[!t]
  \centering
  \includegraphics[width=\linewidth]{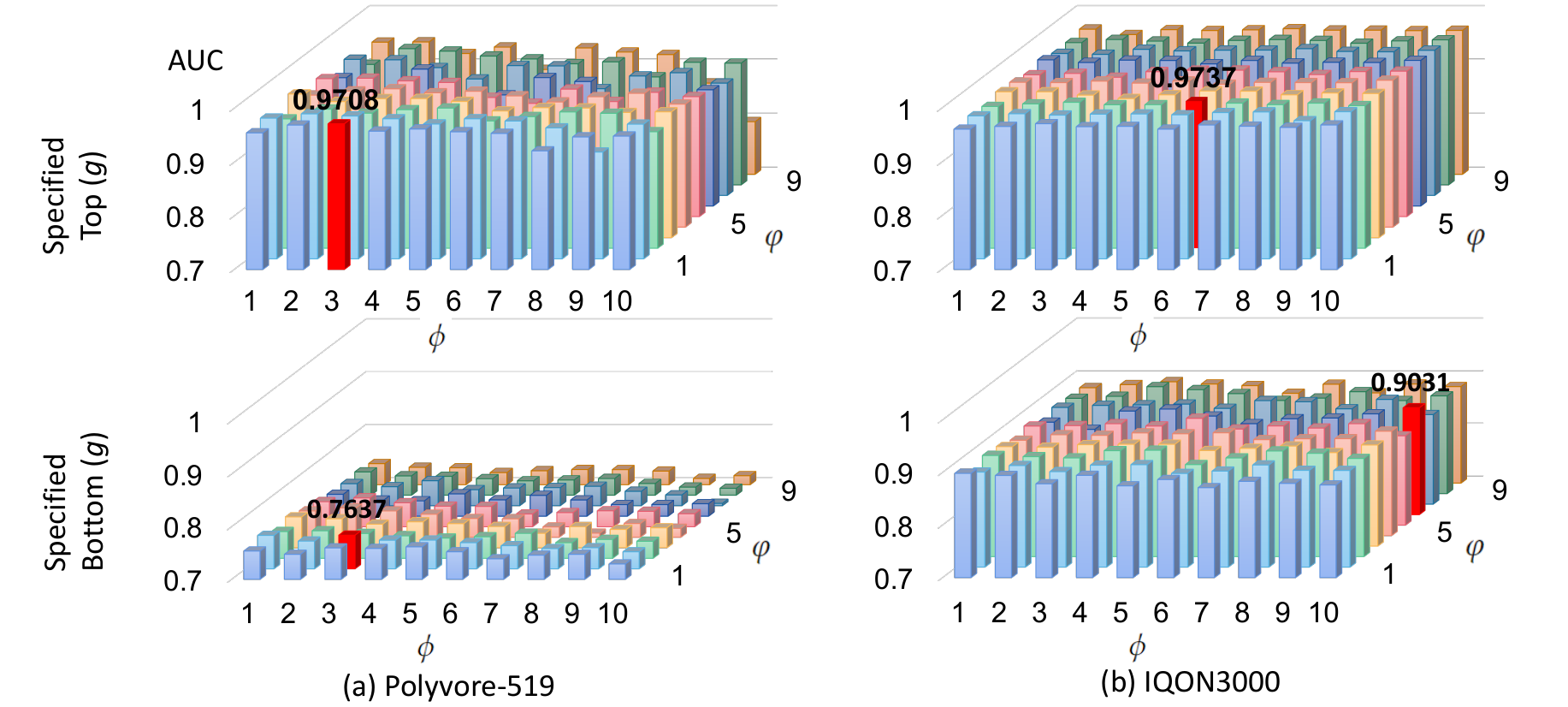}
  \caption{Performance of CR-BPR with different weights for the two consistency regularization branches: (a) Polyvore-519 and (b) IQON3000 datasets.}
  \label{fig.all}
\end{figure}

To get a better understanding of the contribution of each consistency regularization branch, the performance of CR-BPR w/o GC and CR-BPR w/o UC models were assessed with respect to the different weights of $\varphi$ and $\phi$ (eq. (\ref{eq.overall})) for the two datasets. 
As shown in Fig. \ref{fig.all}, the two consistency regularization branches perform similarly irrespective of the changing weights (i.e., importance). Small weight may not be able to produce a consistency regularization effect contributing to the overall performance, whereas large weight will cause the model to be dominated by similar historical choices while largely ignoring user preference and product matching modeling. To ensure CR-BPR's optimal performance, its performance was evaluated when combining two branches with weights varying from one to ten for each branch. In addition, it can be seen that CR-BPR performs better with IQON3000 than with Polyvore-519, and is also less sensitive to changing weights, as shown in the relatively even performance achieved, for both queries using specified top and bottom clothing. 

\subsubsection{\textbf{The Number of Historical Choices ($N$).}}
\begin{figure}[!t]
  \centering
  \includegraphics[width=\linewidth]{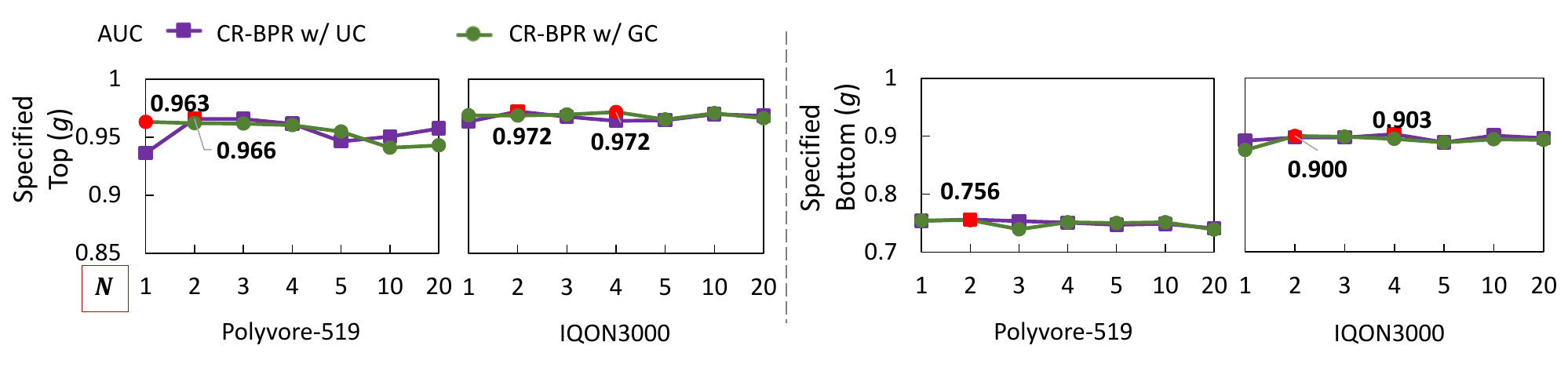}
  \caption{Performance of CR-BPR with GC and UC, respectively, using different number of historical choices $N$ for the (a) Polyvore-519 and (b) IQON3000 datasets.}
  \label{fig.length}
\end{figure}
According to eqs. (\ref{eq.N}) and (\ref{eq.N2}), the number of historical choices $N$ to cover has a
direct impact on the regularization imposed on the user preference and product-matching modeling. The performance curves in Fig.~\ref{fig.length} show the influence of the filtered historical choices on each consistency regularization branch for changing $N$. The historical choice numbers ($N$) were evaluated for a wide range of [1,2,3,4,5,10,20], based on experimental observation and experience. For a valid comparison, the product which is the most similar to the target matching garment is used to fill in the remaining spaces of $gr$ and $ur$ lists, for which there are insufficient information concerning historical choices ($<N$), whereas for cases without historical interaction data, the historical choices were replaced with \textit{popular products} chosen by most users. The results in Fig.~\ref{fig.length} show that the two consistency branches are less sensitive to $N$ for IQON3000 than for Polyvore-519, since the data sparsity problem is more serious for the Polyvore-519 dataset. Furthermore, the best performance mostly lies when $N=2$, this being reasonable since too small a number may overly restrict the current model in terms of learning current preference while overlooking the general preference, whereas too large a number may introduce noise irrelevant to the current matching modeling. In conclusion, reasonable use of previous decisions, regulated by preference and matching consistency, can greatly facilitate and improve personalized clothing matching modeling.

\section{Conclusions and Future Work}\label{sec_5}

To enhance the online shopping experience, personalized fashion recommendation research and applications have gained increasing attention. Personalized complementary clothing recommendation is important in this context, aiming to utilize advanced machine learning techniques to accurately predict and match clothing products based on user preferences, or product features for each specific user. Existing methods have demonstrated the effectiveness in modeling user preference for specified matching product, and the compatibility between given product and matching product.
To further meet the requirements of product matching consistency and user preference consistency, and make the multi-source input data contribute in a balanced manner, a CR-BPR model has been developed in this paper. 
This model leverages two collaborative filtering branches for modeling product matching and user preference, respectively, with a consistency regularization for each branch. In addition, a feature scaling procedure has been utilized to pre-process multi-modal information before model learning. 
Extensive experimental analyses based on two benchmark datasets confirm the effectiveness of the proposed method and the benefits of integrating both product-matching and user preference consistency regularization to recommend complementary clothing products. Performance comparison and qualitative evaluations demonstrate that our proposed method outperforms other existing methods by a large margin.

Despite the proven effectiveness of our CR-BPR model, we will explore further improvements in personalized complementary clothing recommendation as follows. Firstly, our future efforts will focus on developing a more advanced consistency fusion method, such as incorporating consistency knowledge into personalized product matching modeling, to further improve prediction accuracy. Secondly, the current version employs basic technical designs, which can be enhanced with more expressive and sophisticated approaches. Thirdly, the current implementation only considers top and bottom garments for basic technology illustration, necessitating the exploration of fashion coordination with other items or accessories, such as shoes, hats and bags, in future research.

\begin{acks}
The work described in this paper is supported in part by a grant from the Research Grants Council of the Hong Kong Special Administrative Region, China (Grant Number 15602323) and by the Innovation and Technology Commission of Hong Kong, under grant ITP/052/23TP.
		
\end{acks}

\bibliographystyle{ACM-Reference-Format}
\bibliography{base}


\begin{thebibliography}{59}


\ifx \showCODEN    \undefined \def \showCODEN     #1{\unskip}     \fi
\ifx \showDOI      \undefined \def \showDOI       #1{#1}\fi
\ifx \showISBNx    \undefined \def \showISBNx     #1{\unskip}     \fi
\ifx \showISBNxiii \undefined \def \showISBNxiii  #1{\unskip}     \fi
\ifx \showISSN     \undefined \def \showISSN      #1{\unskip}     \fi
\ifx \showLCCN     \undefined \def \showLCCN      #1{\unskip}     \fi
\ifx \shownote     \undefined \def \shownote      #1{#1}          \fi
\ifx \showarticletitle \undefined \def \showarticletitle #1{#1}   \fi
\ifx \showURL      \undefined \def \showURL       {\relax}        \fi
\providecommand\bibfield[2]{#2}
\providecommand\bibinfo[2]{#2}
\providecommand\natexlab[1]{#1}
\providecommand\showeprint[2][]{arXiv:#2}

\bibitem[Chen et~al\mbox{.}(2016)]%
        {chen2016context}
\bibfield{author}{\bibinfo{person}{Tao Chen}, \bibinfo{person}{Xiangnan He}, {and} \bibinfo{person}{Min-Yen Kan}.} \bibinfo{year}{2016}\natexlab{}.
\newblock \showarticletitle{Context-aware image tweet modelling and recommendation}. In \bibinfo{booktitle}{\emph{Proceedings of the 24th ACM international conference on Multimedia}}. \bibinfo{pages}{1018--1027}.
\newblock


\bibitem[Chen et~al\mbox{.}(2019)]%
        {POG}
\bibfield{author}{\bibinfo{person}{Wen Chen}, \bibinfo{person}{Pipei Huang}, \bibinfo{person}{Jiaming Xu}, \bibinfo{person}{Xin Guo}, \bibinfo{person}{Cheng Guo}, \bibinfo{person}{Fei Sun}, \bibinfo{person}{Chao Li}, \bibinfo{person}{Andreas Pfadler}, \bibinfo{person}{Huan Zhao}, {and} \bibinfo{person}{Binqiang Zhao}.} \bibinfo{year}{2019}\natexlab{}.
\newblock \showarticletitle{POG: personalized outfit generation for fashion recommendation at Alibaba iFashion}. In \bibinfo{booktitle}{\emph{Proceedings of the 25th ACM SIGKDD international conference on knowledge discovery \& data mining}}. \bibinfo{pages}{2662--2670}.
\newblock


\bibitem[Ding et~al\mbox{.}(2021)]%
        {ding2021leveraging}
\bibfield{author}{\bibinfo{person}{Yujuan Ding}, \bibinfo{person}{Yunshan Ma}, \bibinfo{person}{Wai~Keung Wong}, {and} \bibinfo{person}{Tat-Seng Chua}.} \bibinfo{year}{2021}\natexlab{}.
\newblock \showarticletitle{Leveraging two types of global graph for sequential fashion recommendation}. In \bibinfo{booktitle}{\emph{Proceedings of the 2021 International Conference on Multimedia Retrieval}}. \bibinfo{pages}{73--81}.
\newblock


\bibitem[Ding et~al\mbox{.}(2023)]%
        {CP}
\bibfield{author}{\bibinfo{person}{Yujuan Ding}, \bibinfo{person}{PY Mok}, \bibinfo{person}{Yi Bin}, \bibinfo{person}{Xun Yang}, {and} \bibinfo{person}{Zhiyong Cheng}.} \bibinfo{year}{2023}\natexlab{}.
\newblock \showarticletitle{Modeling Multi-Relational Connectivity for Personalized Fashion Matching}. In \bibinfo{booktitle}{\emph{Proceedings of the 31st ACM International Conference on Multimedia}}. \bibinfo{pages}{7047--7055}.
\newblock


\bibitem[Dong et~al\mbox{.}(2020)]%
        {dong2020fashion}
\bibfield{author}{\bibinfo{person}{Xue Dong}, \bibinfo{person}{Jianlong Wu}, \bibinfo{person}{Xuemeng Song}, \bibinfo{person}{Hongjun Dai}, {and} \bibinfo{person}{Liqiang Nie}.} \bibinfo{year}{2020}\natexlab{}.
\newblock \showarticletitle{Fashion compatibility modeling through a multi-modal try-on-guided scheme}. In \bibinfo{booktitle}{\emph{Proceedings of the 43rd International ACM SIGIR Conference on Research and Development in Information Retrieval}}. \bibinfo{pages}{771--780}.
\newblock


\bibitem[Feng et~al\mbox{.}(2018)]%
        {part_embedding}
\bibfield{author}{\bibinfo{person}{Zunlei Feng}, \bibinfo{person}{Zhenyun Yu}, \bibinfo{person}{Yezhou Yang}, \bibinfo{person}{Yongcheng Jing}, \bibinfo{person}{Junxiao Jiang}, {and} \bibinfo{person}{Mingli Song}.} \bibinfo{year}{2018}\natexlab{}.
\newblock \showarticletitle{Interpretable partitioned embedding for customized multi-item fashion outfit composition}. In \bibinfo{booktitle}{\emph{Proceedings of the 2018 ACM on International Conference on Multimedia Retrieval}}. \bibinfo{pages}{143--151}.
\newblock


\bibitem[Gao et~al\mbox{.}(2019)]%
        {gao2019fashion}
\bibfield{author}{\bibinfo{person}{Guangyu Gao}, \bibinfo{person}{Liling Liu}, \bibinfo{person}{Li Wang}, {and} \bibinfo{person}{Yihang Zhang}.} \bibinfo{year}{2019}\natexlab{}.
\newblock \showarticletitle{Fashion clothes matching scheme based on Siamese Network and AutoEncoder}.
\newblock \bibinfo{journal}{\emph{Multimedia Systems}} \bibinfo{volume}{25}, \bibinfo{number}{6} (\bibinfo{year}{2019}), \bibinfo{pages}{593--602}.
\newblock


\bibitem[Guan et~al\mbox{.}(2016)]%
        {guan2016apparel}
\bibfield{author}{\bibinfo{person}{Congying Guan}, \bibinfo{person}{Shengfeng Qin}, \bibinfo{person}{Wessie Ling}, {and} \bibinfo{person}{Guofu Ding}.} \bibinfo{year}{2016}\natexlab{}.
\newblock \showarticletitle{Apparel recommendation system evolution: an empirical review}.
\newblock \bibinfo{journal}{\emph{International Journal of Clothing Science and Technology}} (\bibinfo{year}{2016}).
\newblock


\bibitem[Han et~al\mbox{.}(2019)]%
        {han2019prototype}
\bibfield{author}{\bibinfo{person}{Xianjing Han}, \bibinfo{person}{Xuemeng Song}, \bibinfo{person}{Jianhua Yin}, \bibinfo{person}{Yinglong Wang}, {and} \bibinfo{person}{Liqiang Nie}.} \bibinfo{year}{2019}\natexlab{}.
\newblock \showarticletitle{Prototype-guided attribute-wise interpretable scheme for clothing matching}. In \bibinfo{booktitle}{\emph{Proceedings of the 42nd International ACM SIGIR Conference on Research and Development in Information Retrieval}}. \bibinfo{pages}{785--794}.
\newblock


\bibitem[He et~al\mbox{.}(2016)]%
        {resnet50}
\bibfield{author}{\bibinfo{person}{Kaiming He}, \bibinfo{person}{Xiangyu Zhang}, \bibinfo{person}{Shaoqing Ren}, {and} \bibinfo{person}{Jian Sun}.} \bibinfo{year}{2016}\natexlab{}.
\newblock \showarticletitle{Deep residual learning for image recognition}. In \bibinfo{booktitle}{\emph{Proceedings of the IEEE conference on computer vision and pattern recognition}}. \bibinfo{pages}{770--778}.
\newblock


\bibitem[He et~al\mbox{.}(2021)]%
        {he2021click}
\bibfield{author}{\bibinfo{person}{Li He}, \bibinfo{person}{Hongxu Chen}, \bibinfo{person}{Dingxian Wang}, \bibinfo{person}{Shoaib Jameel}, \bibinfo{person}{Philip Yu}, {and} \bibinfo{person}{Guandong Xu}.} \bibinfo{year}{2021}\natexlab{}.
\newblock \showarticletitle{Click-through rate prediction with multi-modal hypergraphs}. In \bibinfo{booktitle}{\emph{Proceedings of the 30th ACM international conference on information \& knowledge management}}. \bibinfo{pages}{690--699}.
\newblock


\bibitem[He and McAuley(2016)]%
        {VBPR}
\bibfield{author}{\bibinfo{person}{Ruining He} {and} \bibinfo{person}{Julian McAuley}.} \bibinfo{year}{2016}\natexlab{}.
\newblock \showarticletitle{VBPR: visual bayesian personalized ranking from implicit feedback}. In \bibinfo{booktitle}{\emph{Proceedings of the AAAI conference on artificial intelligence}}, Vol.~\bibinfo{volume}{30}.
\newblock


\bibitem[He et~al\mbox{.}(2020)]%
        {he2020consistency}
\bibfield{author}{\bibinfo{person}{Yun He}, \bibinfo{person}{Yin Zhang}, \bibinfo{person}{Weiwen Liu}, {and} \bibinfo{person}{James Caverlee}.} \bibinfo{year}{2020}\natexlab{}.
\newblock \showarticletitle{Consistency-aware recommendation for user-generated item list continuation}. In \bibinfo{booktitle}{\emph{Proceedings of the 13th international conference on web search and data mining}}. \bibinfo{pages}{250--258}.
\newblock


\bibitem[Hoerl and Kennard(1970)]%
        {hoerl1970ridge}
\bibfield{author}{\bibinfo{person}{Arthur~E Hoerl} {and} \bibinfo{person}{Robert~W Kennard}.} \bibinfo{year}{1970}\natexlab{}.
\newblock \showarticletitle{Ridge regression: Biased estimation for nonorthogonal problems}.
\newblock \bibinfo{journal}{\emph{Technometrics}} \bibinfo{volume}{12}, \bibinfo{number}{1} (\bibinfo{year}{1970}), \bibinfo{pages}{55--67}.
\newblock


\bibitem[Hou et~al\mbox{.}(2019)]%
        {hou2019explainable}
\bibfield{author}{\bibinfo{person}{Min Hou}, \bibinfo{person}{Le Wu}, \bibinfo{person}{Enhong Chen}, \bibinfo{person}{Zhi Li}, \bibinfo{person}{Vincent~W Zheng}, {and} \bibinfo{person}{Qi Liu}.} \bibinfo{year}{2019}\natexlab{}.
\newblock \showarticletitle{Explainable fashion recommendation: A semantic attribute region guided approach}.
\newblock \bibinfo{journal}{\emph{arXiv preprint arXiv:1905.12862}} (\bibinfo{year}{2019}).
\newblock


\bibitem[Hu et~al\mbox{.}(2008)]%
        {hu2008collaborative}
\bibfield{author}{\bibinfo{person}{Yifan Hu}, \bibinfo{person}{Yehuda Koren}, {and} \bibinfo{person}{Chris Volinsky}.} \bibinfo{year}{2008}\natexlab{}.
\newblock \showarticletitle{Collaborative filtering for implicit feedback datasets}. In \bibinfo{booktitle}{\emph{2008 Eighth IEEE international conference on data mining}}. Ieee, \bibinfo{pages}{263--272}.
\newblock


\bibitem[Jing et~al\mbox{.}(2019)]%
        {jing2019low}
\bibfield{author}{\bibinfo{person}{Peiguang Jing}, \bibinfo{person}{Shu Ye}, \bibinfo{person}{Liqiang Nie}, \bibinfo{person}{Jing Liu}, {and} \bibinfo{person}{Yuting Su}.} \bibinfo{year}{2019}\natexlab{}.
\newblock \showarticletitle{Low-rank regularized multi-representation learning for fashion compatibility prediction}.
\newblock \bibinfo{journal}{\emph{IEEE Transactions on Multimedia}} \bibinfo{volume}{22}, \bibinfo{number}{6} (\bibinfo{year}{2019}), \bibinfo{pages}{1555--1566}.
\newblock


\bibitem[Jing et~al\mbox{.}(2021)]%
        {tmm2022}
\bibfield{author}{\bibinfo{person}{Peiguang Jing}, \bibinfo{person}{Jing Zhang}, \bibinfo{person}{Liqiang Nie}, \bibinfo{person}{Shu Ye}, \bibinfo{person}{Jing Liu}, {and} \bibinfo{person}{Yuting Su}.} \bibinfo{year}{2021}\natexlab{}.
\newblock \showarticletitle{Tripartite graph regularized latent low-rank representation for fashion compatibility prediction}.
\newblock \bibinfo{journal}{\emph{IEEE Transactions on Multimedia}}  \bibinfo{volume}{24} (\bibinfo{year}{2021}), \bibinfo{pages}{1277--1287}.
\newblock


\bibitem[Kaicheng et~al\mbox{.}(2021)]%
        {kaicheng2021modeling}
\bibfield{author}{\bibinfo{person}{Pang Kaicheng}, \bibinfo{person}{Zou Xingxing}, {and} \bibinfo{person}{Wai~Keung Wong}.} \bibinfo{year}{2021}\natexlab{}.
\newblock \showarticletitle{Modeling Fashion Compatibility with Explanation by using Bidirectional LSTM}. In \bibinfo{booktitle}{\emph{Proceedings of the IEEE/CVF Conference on Computer Vision and Pattern Recognition}}. \bibinfo{pages}{3894--3898}.
\newblock


\bibitem[Kim(2009)]%
        {kim2009applying}
\bibfield{author}{\bibinfo{person}{Hyoung-Do Kim}.} \bibinfo{year}{2009}\natexlab{}.
\newblock \showarticletitle{Applying consistency-based trust definition to collaborative filtering}.
\newblock \bibinfo{journal}{\emph{KSII Transactions on Internet and Information Systems (TIIS)}} \bibinfo{volume}{3}, \bibinfo{number}{4} (\bibinfo{year}{2009}), \bibinfo{pages}{366--375}.
\newblock


\bibitem[Kingma and Ba(2014)]%
        {ADAM}
\bibfield{author}{\bibinfo{person}{Diederik~P Kingma} {and} \bibinfo{person}{Jimmy Ba}.} \bibinfo{year}{2014}\natexlab{}.
\newblock \showarticletitle{Adam: A method for stochastic optimization}.
\newblock \bibinfo{journal}{\emph{arXiv preprint arXiv:1412.6980}} (\bibinfo{year}{2014}).
\newblock


\bibitem[Krizhevsky et~al\mbox{.}(2017)]%
        {alexnet}
\bibfield{author}{\bibinfo{person}{Alex Krizhevsky}, \bibinfo{person}{Ilya Sutskever}, {and} \bibinfo{person}{Geoffrey~E Hinton}.} \bibinfo{year}{2017}\natexlab{}.
\newblock \showarticletitle{Imagenet classification with deep convolutional neural networks}.
\newblock \bibinfo{journal}{\emph{Commun. ACM}} \bibinfo{volume}{60}, \bibinfo{number}{6} (\bibinfo{year}{2017}), \bibinfo{pages}{84--90}.
\newblock


\bibitem[Lei et~al\mbox{.}(2023)]%
        {lei2023learning}
\bibfield{author}{\bibinfo{person}{Fei Lei}, \bibinfo{person}{Zhongqi Cao}, \bibinfo{person}{Yuning Yang}, \bibinfo{person}{Yibo Ding}, {and} \bibinfo{person}{Cong Zhang}.} \bibinfo{year}{2023}\natexlab{}.
\newblock \showarticletitle{Learning the user’s deeper preferences for multi-modal recommendation systems}.
\newblock \bibinfo{journal}{\emph{ACM Transactions on Multimedia Computing, Communications and Applications}} \bibinfo{volume}{19}, \bibinfo{number}{3s} (\bibinfo{year}{2023}), \bibinfo{pages}{1--18}.
\newblock


\bibitem[Li et~al\mbox{.}(2021)]%
        {li2021attribute}
\bibfield{author}{\bibinfo{person}{Yang Li}, \bibinfo{person}{Tong Chen}, {and} \bibinfo{person}{Zi Huang}.} \bibinfo{year}{2021}\natexlab{}.
\newblock \showarticletitle{Attribute-aware explainable complementary clothing recommendation}.
\newblock \bibinfo{journal}{\emph{World Wide Web}} \bibinfo{volume}{24}, \bibinfo{number}{5} (\bibinfo{year}{2021}), \bibinfo{pages}{1885--1901}.
\newblock


\bibitem[Li et~al\mbox{.}(2019)]%
        {li2019semi}
\bibfield{author}{\bibinfo{person}{Zekun Li}, \bibinfo{person}{Zeyu Cui}, \bibinfo{person}{Shu Wu}, \bibinfo{person}{Xiaoyu Zhang}, {and} \bibinfo{person}{Liang Wang}.} \bibinfo{year}{2019}\natexlab{}.
\newblock \showarticletitle{Semi-supervised compatibility learning across categories for clothing matching}. In \bibinfo{booktitle}{\emph{2019 IEEE International Conference on Multimedia and Expo (ICME)}}. IEEE, \bibinfo{pages}{484--489}.
\newblock


\bibitem[Liang et~al\mbox{.}(2016)]%
        {liang2016factorization}
\bibfield{author}{\bibinfo{person}{Dawen Liang}, \bibinfo{person}{Jaan Altosaar}, \bibinfo{person}{Laurent Charlin}, {and} \bibinfo{person}{David~M Blei}.} \bibinfo{year}{2016}\natexlab{}.
\newblock \showarticletitle{Factorization meets the item embedding: Regularizing matrix factorization with item co-occurrence}. In \bibinfo{booktitle}{\emph{Proceedings of the 10th ACM conference on recommender systems}}. \bibinfo{pages}{59--66}.
\newblock


\bibitem[Liao et~al\mbox{.}(2023)]%
        {liao}
\bibfield{author}{\bibinfo{person}{Shuiying Liao}, \bibinfo{person}{Yujuan Ding}, {and} \bibinfo{person}{PY Mok}.} \bibinfo{year}{2023}\natexlab{}.
\newblock \showarticletitle{Recommendation of mix-and-match clothing by modeling indirect personal compatibility}. In \bibinfo{booktitle}{\emph{Proceedings of the 2023 ACM International Conference on Multimedia Retrieval}}. \bibinfo{pages}{560--564}.
\newblock


\bibitem[Liu et~al\mbox{.}(2021)]%
        {liu2021bayesian}
\bibfield{author}{\bibinfo{person}{Huafeng Liu}, \bibinfo{person}{Liping Jing}, \bibinfo{person}{Jingxuan Wen}, \bibinfo{person}{Pengyu Xu}, \bibinfo{person}{Jian Yu}, {and} \bibinfo{person}{Michael~K Ng}.} \bibinfo{year}{2021}\natexlab{}.
\newblock \showarticletitle{Bayesian additive matrix approximation for social recommendation}.
\newblock \bibinfo{journal}{\emph{ACM Transactions on Knowledge Discovery from Data (TKDD)}} \bibinfo{volume}{16}, \bibinfo{number}{1} (\bibinfo{year}{2021}), \bibinfo{pages}{1--34}.
\newblock


\bibitem[Liu et~al\mbox{.}(2024b)]%
        {liu2024outfit}
\bibfield{author}{\bibinfo{person}{Hong Liu}, \bibinfo{person}{Li Li}, \bibinfo{person}{Neng Yu}, \bibinfo{person}{Kai Ma}, \bibinfo{person}{Tao Peng}, {and} \bibinfo{person}{Xinrong Hu}.} \bibinfo{year}{2024}\natexlab{b}.
\newblock \showarticletitle{Outfit compatibility model using fully connected self-adjusting graph neural network}.
\newblock \bibinfo{journal}{\emph{The Visual Computer}} (\bibinfo{year}{2024}), \bibinfo{pages}{1--13}.
\newblock


\bibitem[Liu et~al\mbox{.}(2024a)]%
        {liu2024unifying}
\bibfield{author}{\bibinfo{person}{Jinhuan Liu}, \bibinfo{person}{Lei Hou}, \bibinfo{person}{Xu Yu}, \bibinfo{person}{Xuemeng Song}, {and} \bibinfo{person}{Zhaochun Ren}.} \bibinfo{year}{2024}\natexlab{a}.
\newblock \showarticletitle{Unifying heterogeneous and homogeneous relations for personalized compatibility modeling}.
\newblock \bibinfo{journal}{\emph{Knowledge-Based Systems}}  \bibinfo{volume}{290} (\bibinfo{year}{2024}), \bibinfo{pages}{111560}.
\newblock


\bibitem[Liu et~al\mbox{.}(2024c)]%
        {liu2024privacy}
\bibfield{author}{\bibinfo{person}{Jing Liu}, \bibinfo{person}{Litao Shang}, \bibinfo{person}{Yuting Su}, \bibinfo{person}{Weizhi Nie}, \bibinfo{person}{Xin Wen}, {and} \bibinfo{person}{Anan Liu}.} \bibinfo{year}{2024}\natexlab{c}.
\newblock \showarticletitle{Privacy-preserving Multi-source Cross-domain Recommendation Based on Knowledge Graph}.
\newblock \bibinfo{journal}{\emph{ACM Transactions on Multimedia Computing, Communications and Applications}} \bibinfo{volume}{20}, \bibinfo{number}{5} (\bibinfo{year}{2024}), \bibinfo{pages}{1--18}.
\newblock


\bibitem[Liu et~al\mbox{.}(2024d)]%
        {liu2024dual}
\bibfield{author}{\bibinfo{person}{Shenghao Liu}, \bibinfo{person}{Yu Zhang}, \bibinfo{person}{Lingzhi Yi}, \bibinfo{person}{Xianjun Deng}, \bibinfo{person}{Laurence~T Yang}, {and} \bibinfo{person}{Bang Wang}.} \bibinfo{year}{2024}\natexlab{d}.
\newblock \showarticletitle{Dual-side Adversarial Learning based Fair Recommendation for Sensitive Attribute Filtering}.
\newblock \bibinfo{journal}{\emph{ACM Transactions on Knowledge Discovery from Data}} (\bibinfo{year}{2024}).
\newblock


\bibitem[Lu et~al\mbox{.}(2019)]%
        {binary}
\bibfield{author}{\bibinfo{person}{Zhi Lu}, \bibinfo{person}{Yang Hu}, \bibinfo{person}{Yunchao Jiang}, \bibinfo{person}{Yan Chen}, {and} \bibinfo{person}{Bing Zeng}.} \bibinfo{year}{2019}\natexlab{}.
\newblock \showarticletitle{Learning binary code for personalized fashion recommendation}. In \bibinfo{booktitle}{\emph{Proceedings of the IEEE/CVF Conference on Computer Vision and Pattern Recognition}}. \bibinfo{pages}{10562--10570}.
\newblock


\bibitem[Ma et~al\mbox{.}(2024)]%
        {ma2024personalized}
\bibfield{author}{\bibinfo{person}{Jianghong Ma}, \bibinfo{person}{Huiyue Sun}, \bibinfo{person}{Dezhao Yang}, {and} \bibinfo{person}{Haijun Zhang}.} \bibinfo{year}{2024}\natexlab{}.
\newblock \showarticletitle{Personalized Fashion Recommendations for Diverse Body Shapes and Local Preferences with Contrastive Multimodal Cross-Attention Network}.
\newblock \bibinfo{journal}{\emph{ACM Transactions on Intelligent Systems and Technology}} (\bibinfo{year}{2024}).
\newblock


\bibitem[Nie et~al\mbox{.}(2023)]%
        {PCE}
\bibfield{author}{\bibinfo{person}{Xiaozhe Nie}, \bibinfo{person}{Zhijie Xu}, \bibinfo{person}{Jianqin Zhang}, {and} \bibinfo{person}{Yu Tian}.} \bibinfo{year}{2023}\natexlab{}.
\newblock \showarticletitle{Attention-Based Personalized Compatibility Learning for Fashion Matching}.
\newblock \bibinfo{journal}{\emph{Applied Sciences}} \bibinfo{volume}{13}, \bibinfo{number}{17} (\bibinfo{year}{2023}), \bibinfo{pages}{9638}.
\newblock


\bibitem[Rendle et~al\mbox{.}(2012)]%
        {BPR}
\bibfield{author}{\bibinfo{person}{Steffen Rendle}, \bibinfo{person}{Christoph Freudenthaler}, \bibinfo{person}{Zeno Gantner}, {and} \bibinfo{person}{Lars Schmidt-Thieme}.} \bibinfo{year}{2012}\natexlab{}.
\newblock \showarticletitle{BPR: Bayesian personalized ranking from implicit feedback}.
\newblock \bibinfo{journal}{\emph{arXiv preprint arXiv:1205.2618}} (\bibinfo{year}{2012}).
\newblock


\bibitem[Sagar et~al\mbox{.}(2020)]%
        {PAI}
\bibfield{author}{\bibinfo{person}{Dikshant Sagar}, \bibinfo{person}{Jatin Garg}, \bibinfo{person}{Prarthana Kansal}, \bibinfo{person}{Sejal Bhalla}, \bibinfo{person}{Rajiv~Ratn Shah}, {and} \bibinfo{person}{Yi Yu}.} \bibinfo{year}{2020}\natexlab{}.
\newblock \showarticletitle{Pai-bpr: Personalized outfit recommendation scheme with attribute-wise interpretability}. In \bibinfo{booktitle}{\emph{2020 IEEE Sixth International Conference on Multimedia Big Data (BigMM)}}. IEEE, \bibinfo{pages}{221--230}.
\newblock


\bibitem[Sheth et~al\mbox{.}(2023)]%
        {sheth2023causal}
\bibfield{author}{\bibinfo{person}{Paras Sheth}, \bibinfo{person}{Ruocheng Guo}, \bibinfo{person}{Lu Cheng}, \bibinfo{person}{Huan Liu}, {and} \bibinfo{person}{Kasim~Sel{\c{c}}uk Candan}.} \bibinfo{year}{2023}\natexlab{}.
\newblock \showarticletitle{Causal disentanglement for implicit recommendations with network information}.
\newblock \bibinfo{journal}{\emph{ACM Transactions on Knowledge Discovery from Data}} \bibinfo{volume}{17}, \bibinfo{number}{7} (\bibinfo{year}{2023}), \bibinfo{pages}{1--18}.
\newblock


\bibitem[Song et~al\mbox{.}(2017)]%
        {song2017neurostylist}
\bibfield{author}{\bibinfo{person}{Xuemeng Song}, \bibinfo{person}{Fuli Feng}, \bibinfo{person}{Jinhuan Liu}, \bibinfo{person}{Zekun Li}, \bibinfo{person}{Liqiang Nie}, {and} \bibinfo{person}{Jun Ma}.} \bibinfo{year}{2017}\natexlab{}.
\newblock \showarticletitle{Neurostylist: Neural compatibility modeling for clothing matching}. In \bibinfo{booktitle}{\emph{Proceedings of the 25th ACM international conference on Multimedia}}. \bibinfo{pages}{753--761}.
\newblock


\bibitem[Song et~al\mbox{.}(2019)]%
        {GPBPR}
\bibfield{author}{\bibinfo{person}{Xuemeng Song}, \bibinfo{person}{Xianjing Han}, \bibinfo{person}{Yunkai Li}, \bibinfo{person}{Jingyuan Chen}, \bibinfo{person}{Xin-Shun Xu}, {and} \bibinfo{person}{Liqiang Nie}.} \bibinfo{year}{2019}\natexlab{}.
\newblock \showarticletitle{GP-BPR: Personalized compatibility modeling for clothing matching}. In \bibinfo{booktitle}{\emph{Proceedings of the 27th ACM International Conference on Multimedia}}. \bibinfo{pages}{320--328}.
\newblock


\bibitem[Sun et~al\mbox{.}(2018)]%
        {style_consistency}
\bibfield{author}{\bibinfo{person}{Guang-Lu Sun}, \bibinfo{person}{Zhi-Qi Cheng}, \bibinfo{person}{Xiao Wu}, {and} \bibinfo{person}{Qiang Peng}.} \bibinfo{year}{2018}\natexlab{}.
\newblock \showarticletitle{Personalized clothing recommendation combining user social circle and fashion style consistency}.
\newblock \bibinfo{journal}{\emph{Multimedia Tools and Applications}} \bibinfo{volume}{77}, \bibinfo{number}{14} (\bibinfo{year}{2018}), \bibinfo{pages}{17731--17754}.
\newblock


\bibitem[Sun et~al\mbox{.}(2020b)]%
        {sun2020learning}
\bibfield{author}{\bibinfo{person}{Guang-Lu Sun}, \bibinfo{person}{Jun-Yan He}, \bibinfo{person}{Xiao Wu}, \bibinfo{person}{Bo Zhao}, {and} \bibinfo{person}{Qiang Peng}.} \bibinfo{year}{2020}\natexlab{b}.
\newblock \showarticletitle{Learning fashion compatibility across categories with deep multimodal neural networks}.
\newblock \bibinfo{journal}{\emph{Neurocomputing}}  \bibinfo{volume}{395} (\bibinfo{year}{2020}), \bibinfo{pages}{237--246}.
\newblock


\bibitem[Sun et~al\mbox{.}(2020a)]%
        {sun2020multi}
\bibfield{author}{\bibinfo{person}{Rui Sun}, \bibinfo{person}{Xuezhi Cao}, \bibinfo{person}{Yan Zhao}, \bibinfo{person}{Junchen Wan}, \bibinfo{person}{Kun Zhou}, \bibinfo{person}{Fuzheng Zhang}, \bibinfo{person}{Zhongyuan Wang}, {and} \bibinfo{person}{Kai Zheng}.} \bibinfo{year}{2020}\natexlab{a}.
\newblock \showarticletitle{Multi-modal knowledge graphs for recommender systems}. In \bibinfo{booktitle}{\emph{Proceedings of the 29th ACM international conference on information \& knowledge management}}. \bibinfo{pages}{1405--1414}.
\newblock


\bibitem[Tibshirani(1996)]%
        {tibshirani1996regression}
\bibfield{author}{\bibinfo{person}{Robert Tibshirani}.} \bibinfo{year}{1996}\natexlab{}.
\newblock \showarticletitle{Regression shrinkage and selection via the lasso}.
\newblock \bibinfo{journal}{\emph{Journal of the Royal Statistical Society: Series B (Methodological)}} \bibinfo{volume}{58}, \bibinfo{number}{1} (\bibinfo{year}{1996}), \bibinfo{pages}{267--288}.
\newblock


\bibitem[Trakulwaranont et~al\mbox{.}(2022)]%
        {trakulwaranont2022personalized}
\bibfield{author}{\bibinfo{person}{Donnaphat Trakulwaranont}, \bibinfo{person}{Marc~A Kastner}, {and} \bibinfo{person}{Shin’ichi Satoh}.} \bibinfo{year}{2022}\natexlab{}.
\newblock \showarticletitle{Personalized Fashion Recommendation Using Pairwise Attention}. In \bibinfo{booktitle}{\emph{International Conference on Multimedia Modeling}}. Springer, \bibinfo{pages}{218--229}.
\newblock


\bibitem[Tran et~al\mbox{.}(2018)]%
        {tran2018regularizing}
\bibfield{author}{\bibinfo{person}{Thanh Tran}, \bibinfo{person}{Kyumin Lee}, \bibinfo{person}{Yiming Liao}, {and} \bibinfo{person}{Dongwon Lee}.} \bibinfo{year}{2018}\natexlab{}.
\newblock \showarticletitle{Regularizing matrix factorization with user and item embeddings for recommendation}. In \bibinfo{booktitle}{\emph{Proceedings of the 27th ACM international conference on information and knowledge management}}. \bibinfo{pages}{687--696}.
\newblock


\bibitem[Vasileva et~al\mbox{.}(2018)]%
        {vasileva2018learning}
\bibfield{author}{\bibinfo{person}{Mariya~I Vasileva}, \bibinfo{person}{Bryan~A Plummer}, \bibinfo{person}{Krishna Dusad}, \bibinfo{person}{Shreya Rajpal}, \bibinfo{person}{Ranjitha Kumar}, {and} \bibinfo{person}{David Forsyth}.} \bibinfo{year}{2018}\natexlab{}.
\newblock \showarticletitle{Learning type-aware embeddings for fashion compatibility}. In \bibinfo{booktitle}{\emph{Proceedings of the European conference on computer vision (ECCV)}}. \bibinfo{pages}{390--405}.
\newblock


\bibitem[Veli{\v{c}}kovi{\'c} et~al\mbox{.}(2017)]%
        {velivckovic2017graph}
\bibfield{author}{\bibinfo{person}{Petar Veli{\v{c}}kovi{\'c}}, \bibinfo{person}{Guillem Cucurull}, \bibinfo{person}{Arantxa Casanova}, \bibinfo{person}{Adriana Romero}, \bibinfo{person}{Pietro Lio}, {and} \bibinfo{person}{Yoshua Bengio}.} \bibinfo{year}{2017}\natexlab{}.
\newblock \showarticletitle{Graph attention networks}.
\newblock \bibinfo{journal}{\emph{arXiv preprint arXiv:1710.10903}} (\bibinfo{year}{2017}).
\newblock


\bibitem[Wang et~al\mbox{.}(2023)]%
        {wang2023multifaceted}
\bibfield{author}{\bibinfo{person}{Chunyang Wang}, \bibinfo{person}{Yanmin Zhu}, \bibinfo{person}{Haobing Liu}, \bibinfo{person}{Tianzi Zang}, \bibinfo{person}{Ke Wang}, {and} \bibinfo{person}{Jiadi Yu}.} \bibinfo{year}{2023}\natexlab{}.
\newblock \showarticletitle{Multifaceted Relation-aware Meta-learning with Dual Customization for User Cold-start Recommendation}.
\newblock \bibinfo{journal}{\emph{ACM Transactions on Knowledge Discovery from Data}} \bibinfo{volume}{17}, \bibinfo{number}{9} (\bibinfo{year}{2023}), \bibinfo{pages}{1--27}.
\newblock


\bibitem[Wang et~al\mbox{.}(2017)]%
        {wang2017item}
\bibfield{author}{\bibinfo{person}{Xiang Wang}, \bibinfo{person}{Xiangnan He}, \bibinfo{person}{Liqiang Nie}, {and} \bibinfo{person}{Tat-Seng Chua}.} \bibinfo{year}{2017}\natexlab{}.
\newblock \showarticletitle{Item silk road: Recommending items from information domains to social users}. In \bibinfo{booktitle}{\emph{Proceedings of the 40th International ACM SIGIR conference on Research and Development in Information Retrieval}}. \bibinfo{pages}{185--194}.
\newblock


\bibitem[Xu et~al\mbox{.}(2020)]%
        {xu2020neural}
\bibfield{author}{\bibinfo{person}{Yuanbo Xu}, \bibinfo{person}{Yongjian Yang}, \bibinfo{person}{En Wang}, \bibinfo{person}{Jiayu Han}, \bibinfo{person}{Fuzhen Zhuang}, \bibinfo{person}{Zhiwen Yu}, {and} \bibinfo{person}{Hui Xiong}.} \bibinfo{year}{2020}\natexlab{}.
\newblock \showarticletitle{Neural serendipity recommendation: Exploring the balance between accuracy and novelty with sparse explicit feedback}.
\newblock \bibinfo{journal}{\emph{ACM Transactions on Knowledge Discovery from Data (TKDD)}} \bibinfo{volume}{14}, \bibinfo{number}{4} (\bibinfo{year}{2020}), \bibinfo{pages}{1--25}.
\newblock


\bibitem[Yan et~al\mbox{.}(2022)]%
        {computer}
\bibfield{author}{\bibinfo{person}{An Yan}, \bibinfo{person}{Chaosheng Dong}, \bibinfo{person}{Yan Gao}, \bibinfo{person}{Jinmiao Fu}, \bibinfo{person}{Tong Zhao}, \bibinfo{person}{Yi Sun}, {and} \bibinfo{person}{Julian McAuley}.} \bibinfo{year}{2022}\natexlab{}.
\newblock \showarticletitle{Personalized complementary product recommendation}. In \bibinfo{booktitle}{\emph{Companion Proceedings of the Web Conference 2022}}. \bibinfo{pages}{146--151}.
\newblock


\bibitem[Yang et~al\mbox{.}(2020)]%
        {yang2020learning}
\bibfield{author}{\bibinfo{person}{Xun Yang}, \bibinfo{person}{Xiaoyu Du}, {and} \bibinfo{person}{Meng Wang}.} \bibinfo{year}{2020}\natexlab{}.
\newblock \showarticletitle{Learning to match on graph for fashion compatibility modeling}. In \bibinfo{booktitle}{\emph{Proceedings of the AAAI Conference on Artificial Intelligence}}, Vol.~\bibinfo{volume}{34}. \bibinfo{pages}{287--294}.
\newblock


\bibitem[Yujuan et~al\mbox{.}(2022)]%
        {9451610}
\bibfield{author}{\bibinfo{person}{Ding Yujuan}, \bibinfo{person}{Ma Yunshan}, \bibinfo{person}{Wai~Keung Wong}, {and} \bibinfo{person}{Tat-Seng Chua}.} \bibinfo{year}{2022}\natexlab{}.
\newblock \showarticletitle{Modeling Instant User Intent and Content-Level Transition for Sequential Fashion Recommendation}.
\newblock \bibinfo{journal}{\emph{IEEE Transactions on Multimedia}}  \bibinfo{volume}{24} (\bibinfo{year}{2022}), \bibinfo{pages}{2687--2700}.
\newblock
\urldef\tempurl%
\url{https://doi.org/10.1109/TMM.2021.3088281}
\showDOI{\tempurl}


\bibitem[Zhan et~al\mbox{.}(2021)]%
        {A3}
\bibfield{author}{\bibinfo{person}{Huijing Zhan}, \bibinfo{person}{Jie Lin}, \bibinfo{person}{Kenan~Emir Ak}, \bibinfo{person}{Boxin Shi}, \bibinfo{person}{Ling-Yu Duan}, {and} \bibinfo{person}{Alex~C Kot}.} \bibinfo{year}{2021}\natexlab{}.
\newblock \showarticletitle{$A^{3}$-FKG: Attentive Attribute-Aware Fashion Knowledge Graph for Outfit Preference Prediction}.
\newblock \bibinfo{journal}{\emph{IEEE Transactions on Multimedia}}  \bibinfo{volume}{24} (\bibinfo{year}{2021}), \bibinfo{pages}{819--831}.
\newblock


\bibitem[Zhang et~al\mbox{.}(2013)]%
        {AUC}
\bibfield{author}{\bibinfo{person}{Hanwang Zhang}, \bibinfo{person}{Zheng-Jun Zha}, \bibinfo{person}{Yang Yang}, \bibinfo{person}{Shuicheng Yan}, \bibinfo{person}{Yue Gao}, {and} \bibinfo{person}{Tat-Seng Chua}.} \bibinfo{year}{2013}\natexlab{}.
\newblock \showarticletitle{Attribute-augmented semantic hierarchy: towards bridging semantic gap and intention gap in image retrieval}. In \bibinfo{booktitle}{\emph{Proceedings of the 21st ACM international conference on Multimedia}}. \bibinfo{pages}{33--42}.
\newblock


\bibitem[Zhao et~al\mbox{.}(2024)]%
        {zhao2024hierarchical}
\bibfield{author}{\bibinfo{person}{Hong Zhao}, \bibinfo{person}{Zhengyu Li}, \bibinfo{person}{Wenwei He}, {and} \bibinfo{person}{Yan Zhao}.} \bibinfo{year}{2024}\natexlab{}.
\newblock \showarticletitle{Hierarchical Convolutional Neural Network with Knowledge Complementation for Long-Tailed Classification}.
\newblock \bibinfo{journal}{\emph{ACM Transactions on Knowledge Discovery from Data}} \bibinfo{volume}{18}, \bibinfo{number}{6} (\bibinfo{year}{2024}), \bibinfo{pages}{1--22}.
\newblock


\bibitem[Zhu et~al\mbox{.}(2019)]%
        {zhu2019dtcdr}
\bibfield{author}{\bibinfo{person}{Feng Zhu}, \bibinfo{person}{Chaochao Chen}, \bibinfo{person}{Yan Wang}, \bibinfo{person}{Guanfeng Liu}, {and} \bibinfo{person}{Xiaolin Zheng}.} \bibinfo{year}{2019}\natexlab{}.
\newblock \showarticletitle{Dtcdr: A framework for dual-target cross-domain recommendation}. In \bibinfo{booktitle}{\emph{Proceedings of the 28th ACM international conference on information and knowledge management}}. \bibinfo{pages}{1533--1542}.
\newblock


\bibitem[Zhu et~al\mbox{.}(2017)]%
        {broadlearning}
\bibfield{author}{\bibinfo{person}{Junxing Zhu}, \bibinfo{person}{Jiawei Zhang}, \bibinfo{person}{Lifang He}, \bibinfo{person}{Quanyuan Wu}, \bibinfo{person}{Bin Zhou}, \bibinfo{person}{Chenwei Zhang}, {and} \bibinfo{person}{Philip~S Yu}.} \bibinfo{year}{2017}\natexlab{}.
\newblock \showarticletitle{Broad learning based multi-source collaborative recommendation}. In \bibinfo{booktitle}{\emph{Proceedings of the 2017 ACM on Conference on Information and Knowledge Management}}. \bibinfo{pages}{1409--1418}.
\newblock


\end{thebibliography}


\end{document}